%% file: main.tex
\documentclass[twocolumn]{aastex631}

\input{custom_commands}

\begin{document}

\title{ML in Astrophysical Turbulence I: Predicting Prestellar Cores in \\ Magnetized Molecular Clouds using eXtreme Gradient Boosting}

\author[0000-0002-7557-1890]{Nikhil P. S. Bisht}
\affiliation{Department of Physics, Florida State University, Tallahassee, Florida, USA}

\author[0000-0001-6661-2243]{David C. Collins}
\affiliation{Department of Physics, Florida State University, Tallahassee, Florida, USA}



\input{0_abstract}

\input{1_Introduction}

\input{2_Sim_and_Data}

\input{3_Methods}

\input{4_Results}

\input{5_Discussion}

\input{6_Conclusion}

\input{7_Acknowledgements}

\bibliography{bibliography}{}
\bibliographystyle{aasjournal}



\end{document}

%% file: custom_commands.tex
\usepackage{graphicx, natbib, color, bm, amsmath, epsfig, hyperref}

\newcommand{\K}{\,\rm K}

\newcommand{\p}[1]{#1^\prime }

\definecolor{orange}{rgb}{1.        ,  0.54,  0}

\definecolor{meta}{rgb}{0.371,0.617,0.625} 

\newcommand{\dc}[1]{}

\definecolor{pink}{rgb}{1.        ,  0.75294118,  0.79607843}
\definecolor{maroon}{rgb}{0.69019608,  0.18823529,  0.37647059}

\definecolor{gray}{rgb}{0.5,0.5,0.5}

%% file: 0_abstract.tex
\begin{abstract}
Giant Molecular Clouds (GMCs) are dominated by supersonic turbulence and magnetic fields, creating a complex network of shocks and filaments that regulate star formation, making it challenging to predict which individual gas parcels will collapse into star-forming cores. Here, we present an XGBoost machine learning framework to forecast the 3D Lagrangian trajectories and gravitational collapse fate of prestellar gas. Models were trained on $\sim 2.1$ million tracer particles from self-gravitating, turbulent magnetohydrodynamic (MHD) simulations. By mapping instantaneous point-wise states (position, velocity, density, and magnetic fields) to future coordinates, our regression model predicts continuous 3D spatial evolution over $0.25~t_{\rm ff}$ ($\approx 0.31\text{ Myr}$) with $R^2 > 0.99$. Concurrently, our classifier accurately ranks gravitational core susceptibility ($\text{AUC} = 0.975$). Using SHapley Additive exPlanations (SHAP), we interpret the model's decision hierarchy, revealing that local density, kinetic energy, and magnetic field magnitude dominate core identification. SHAP dependence analysis shows that the model re-derives a sharp non-linear collapse threshold at $\log_{10}(\rho/\rho_0) \approx 0.8$ above which magnetic fields synergistically amplify collapse predictions. Crucially, the classification task reveals a systematically conservative false-positive rate (37\% precision) at absolute binary thresholds. We demonstrate that while point-wise ML models excel at kinematic advection, they mathematically lack the multi-point spatial context required to evaluate local differential operators (such as magnetic directional derivatives, thermal pressure gradients, or field curvature) that are physically necessary to stop collapse. This data-driven approach offers a computationally efficient foundation for advective subgrid modeling while highlighting the necessity of spatial architectures (e.g., Deep Learning) to establish absolute core-formation criteria.
\end{abstract}

\keywords{Molecular clouds (1072) --- Regression (1914) --- Classification  (1907) --- Magnetohydrodynamical simulations (1966)}

%% file: 1_Introduction.tex
\section{Introduction} \label{sec:intro}

The formation of stars is the fundamental driver of galaxy evolution, regulating the interstellar medium (ISM) through feedback and enriching the universe with heavy elements. Despite its importance, the precise mechanism that regulates the rate of star formation remains one of the central problems in astrophysics. Observations indicate that star formation is surprisingly inefficient; only $1\%-2\%$ of a molecular cloud's mass is converted into stars per free-fall time ($\epsilon_{\rm ff} \approx 0.01$) \citep{Krumholz2005, Krumholz2007}. This inefficiency is generally attributed to the interplay between self-gravity, which promotes collapse, and supersonic turbulence and magnetic fields, which provide support and shear apart potential star-forming regions \citep{McKee2007, Federrath2012}.

Giant Molecular Clouds (GMCs) are the primary nurseries for this process. Spanning scales from tens of parsecs ($\sim 10^{19}$ m) down to the size of prestellar cores at about $0.1$ parsecs ($\sim 3 \times 10^{15}$ m), these clouds represent a chaotic environment where gas spans orders of magnitude in density and temperature. The standard paradigm suggests that supersonic turbulence creates a complex network of shocks and filaments \citep{Larson1981, MacLow2004}. Within this turbulent cascade, transient overdensities form and dissolve, but only those fluctuations that are gravitationally bound and magnetically supercritical (a rare subset) succeed in collapsing to form prestellar cores \citep{Shu1977, Padoan2002, Crutcher2012}. 

Once these cores collapse and protostars ignite, they inject significant energy back into their environment. Radiative feedback (e.g., UV heating and photoionization) and protostellar outflows play a critical role in regulating the global star formation efficiency by dispersing the surrounding gas and halting further accretion \citep{McKee2007}. However, as this study focuses strictly on the kinematic origins and the initial turbulent fragmentation of the \textit{prestellar} phase which is prior to the onset of stellar ignition, we approximate the gas thermodynamics using an isothermal equation of state and intentionally omit radiative feedback mechanisms. This isolates the purely gravity-driven turbulence and advective rules our machine learning model aims to learn.

\begin{figure*}[]
\includegraphics[width=2\columnwidth]{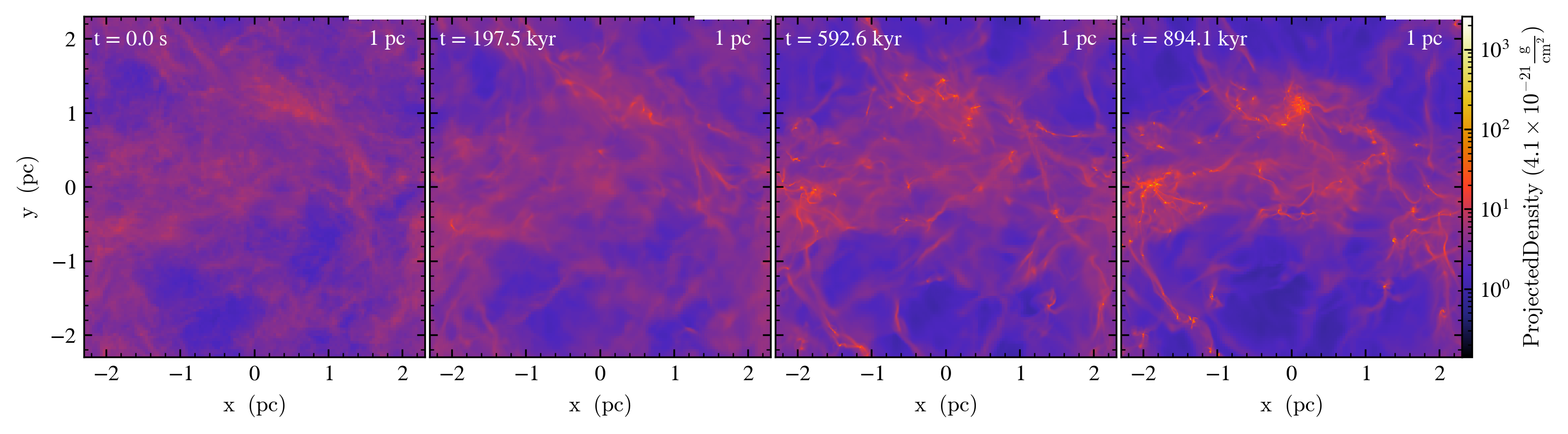}
\caption{Evolution of the turbulent density field. We show the column density projection ($\Sigma$) of the simulation box at four evolutionary stages, ranging from the initial turbulent driving phase ($t=0$) to the onset of widespread core collapse ($t \approx t_{\rm ff}$).
\label{fig:sim}}
\end{figure*}

Observational surveys using tracer molecules (e.g., $^{12}\text{CO}$, $\text{NH}_3$, $\text{N}_2\text{H}^+$) and dust continuum emission have provided a wealth of data regarding the morphology of these regions \citep{Alatalo2013, Evans2009}. Recent high-resolution campaigns with ALMA and observational analysis of cloud statistics \citep{Chevance2022, Sahu2021} have highlighted the environmental dependence of core formation. However, observations suffer from projection effects (2D snapshots of 3D structures) and a lack of temporal evolution. We observe the "before" (diffuse gas) and "after" (protostars), but the critical dynamic phase of collapse is often obscured or ambiguous.

To bridge this gap, high-resolution Magnetohydrodynamic (MHD) simulations have become indispensable. Codes such as \textsc{Enzo}, \textsc{Flash}, and \textsc{Arepo} allow us to model the non-linear equations of fluid dynamics coupled with gravity and magnetic fields \citep{Bryan2014, Fryxell2000, Weinberger2020}. While simulations provide full 6D phase-space information, identifying the Lagrangian history of star-forming gas, i.e., tracing the specific parcels of gas that will eventually become stars, remains a non-trivial data analysis challenge. Traditional methods often rely on thresholding (e.g., density cutoffs) or sink particles \citep{Bate1995, Federrath2010b, Wang2009}, which identify collapse only \textit{after} it has become irreversible.

In recent years, Machine Learning (ML) has emerged as a powerful tool for analyzing the complex, high-dimensional datasets produced by astrophysical simulations. ML techniques have been successfully applied to estimate halo masses \citep{Ntampaka2015}, classify galaxy morphologies \citep{Dieleman2015}, and infer physical parameters from turbulent fields \citep{Peek2019, Salim2024, Schmaltz2025, Zhang2026}. In the context of star formation, recent works have begun to utilize Deep Learning to classify young stellar objects \citep{Tan2024} and identify structures in molecular line data. However, most efforts focus on Convolutional Neural Networks (CNNs) applied to image data.

In this work, we present a novel approach focusing on the \textit{Lagrangian} dynamics of the gas. We frame the evolution of prestellar gas not as a macroscopic image recognition task, but as two distinct tabular supervised learning problems: a continuous regression task to forecast 3D particle advection, and a binary classification task to predict the ultimate accretion fate of the gas. We utilize \textbf{XGBoost} (Extreme Gradient Boosting) \footnote{https://github.com/dmlc/xgboost}, a decision-tree-based ensemble method \citep{Chen2016}, to analyze the trajectories of passive tracer particles in strongly magnetized ($\beta=0.2$) and moderately magnetized ($\beta=2.0$) turbulent MHD simulations. Unlike "black box" neural networks, tree-based models are highly effective for tabular physical data (incorporating density, kinematics, and localized magnetic vectors) and offer strict physical interpretability when combined with feature attribution methods like SHAP.

Our goal is to answer a specific twofold question: \textit{Given the instantaneous, point-wise state of a gas parcel in a turbulent cloud, can we accurately forecast its continuous short-term trajectory and probabilistically determine its ultimate collapse fate?} Furthermore, by explicitly evaluating the limits of what a purely point-wise model can learn, we aim to mathematically demonstrate the necessity of continuous spatial gradients (such as magnetic tension) in regulating gravitational instability.

This paper is structured as follows. Section \ref{sec:sim_n_data} details the MHD simulation setup, the tracer particle methodology, and the data pipeline. Section \ref{sec:methods} describes the problem formulation, feature engineering, the XGBoost architecture, and the SHAP feature attribution framework. Section \ref{sec:results} presents the continuous trajectory forecasting, temporal generalization tests, and the probabilistic limits exposed by our classification task. We discuss the physical implications of our findings regarding magnetic fields, feature importance, and subgrid modeling in Section \ref{sec:discussion}, and conclude in Section \ref{sec:conclusion}.

%% file: 2_Sim_and_Data.tex
\section{Simulations and Data} \label{sec:sim_n_data}

\subsection{Turbulence Box Simulations} \label{sec:sim}
We utilize the open-source adaptive mesh refinement (AMR) code \textsc{Enzo} \citep{Bryan2014} with the constrained transport magnetohydrodynamics (MHD) module \citep{Collins2010}. The code adaptively increases resolution as the collapse proceeds, triggering refinement whenever the local Jeans length, $L_{\rm{J}}=c_s (\pi / G \rho)^{1/2}$, is resolved by fewer than 16 zones. The primary solver employs a higher-order Godunov method, utilizing the linear reconstruction method of \citet{Li2008}, the HLLD Riemann solver of \citet{Mignone2007}, and the electric field correction of \citet{Gardiner2005} to maintain the divergence-free constraint of the magnetic field ($\nabla \cdot \mathbf{B} = 0$).

To track the gas parcels in a Lagrangian framework, we include passive tracer particles that follow the flow via a piecewise linear interpolation of the grid velocity field \citep[see][]{Bryan2014, Collins2023}. 

Initial conditions are generated following the procedure of \citet{MacLow1999}: we initialize a periodic turbulent box with uniform density $\rho_0$ and uniform magnetic field $B_0$. We inject kinetic energy at large scales (wavenumbers between 1 and 2), driving solenoidal velocity patterns, allowing it to cascade to smaller scales via non-linear interactions until a statistically steady state is achieved. At this point, we activate self-gravity, insert uniformly spaced tracer particles, and allow the gas to collapse. 

The simulation is characterized by the sonic Mach number $\mathcal{M}$, the virial parameter $\alpha_{\rm vir}$, and the plasma $\beta_0$:
\begin{align}
\mathcal{M} &= \frac{v_{\rm rms}}{c_s} = 9\\
\alpha_{\rm vir} &= \frac{5 v_{\rm rms}^2}{3 G \rho_0 L_0} = 1\\
\beta_0 &= \frac{8 \pi c_s^2 \rho_0}{B_0^2} = 0.2
\end{align}
where $v_{\rm rms}$ is the root-mean-square velocity, $c_s$ is the sound speed, and $L_0$ is the box size. The magnetic field is initially uniform in the $\hat{x}$ direction. We follow the scaling laid out in \citet{Collins2012}; the length scale of the box is 4.6 pc, the density is 1000 $cm^{-3}$ (or $4.1 \times 10^{-21} g/cm^{-3}$), the time scale is 1.1 Myr, the mass of the box is 5900 $M_{\odot}$, and the magnetic field strength is 13 $\mu G$. The gas is treated as isothermal with a temperature of 10 K. 

The simulation evolves for approximately one free-fall time, $t_{\rm ff}=\sqrt{3 \pi / 32 G \rho_0}$. We initialize $128^3$ tracer particles (one per root-grid zone), totaling $2,097,152$ particles. Tracers are evolved using a second-order predictor-corrector scheme where the particle position is updated as $\mathbf{x}^{n+1} = \mathbf{x}^{n} + \Delta t \, \mathbf{v}_{\rm{p}}$. Refinement continues for 4 levels, yielding a maximum effective resolution of $2048^3$. Further details on the simulation suite can be found in \citep{Collins2023, Collins2024}.

\begin{figure*}[]
\includegraphics[width=2\columnwidth]{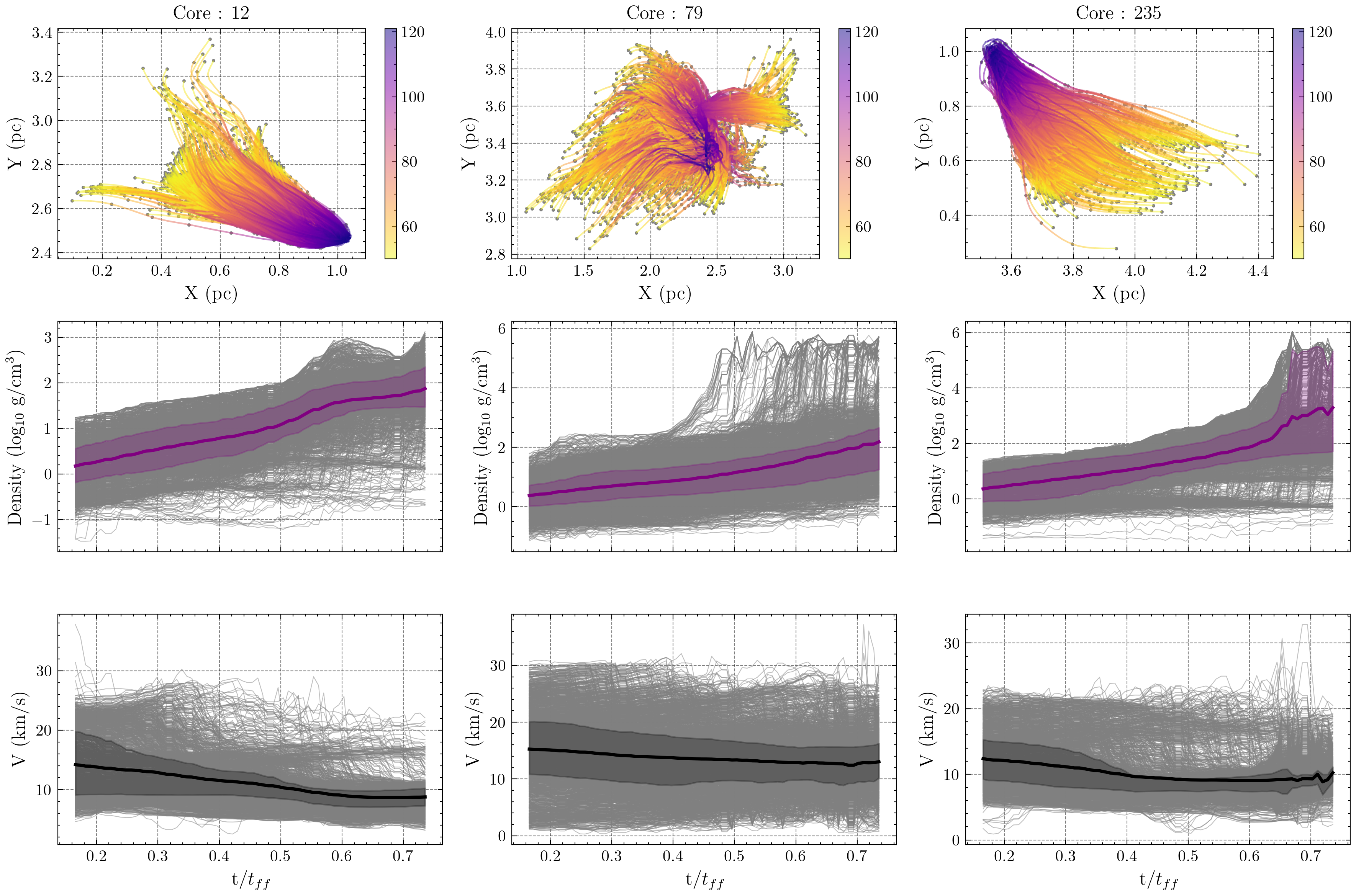}
\caption{Evolution of three representative prestellar cores. \textbf{Top:} Projected trajectories of tracer particles belonging to each core from $t=0.5 t_{\rm ff}$ (yellow) to $t=1.0 t_{\rm ff}$ (purple). \textbf{Middle:} Evolution of the local gas density surrounding the tracers. \textbf{Bottom:} Evolution of the velocity magnitude. The solid lines represent the median value for the core's constituent particles, while the shaded regions indicate the 25th-75th percentile range. As collapse proceeds, density increases rapidly while the velocity dispersion decreases, indicating coherent infall.
\label{fig:tracks}}
\end{figure*}

\subsection{Data Extraction and Preparation} \label{sec:data_ex_prep}

The simulation state is archived at high frequency, resulting in 121 snapshots spanning $1.0~t_{\rm ff}$. Each snapshot represents a physical time interval of approximately $0.01$ Myr. For every snapshot $i$, we extract the full phase-space state for each tracer particle: position $\mathbf{x}_i$, velocity $\mathbf{v}_i$, local magnetic field $\mathbf{B}_i = (B_{x,i}, B_{y,i}, B_{z,i})$, and local gas density $\rho_i$. This collection constitutes our \emph{Raw Simulation Dataset}.

As the turbulence decays and gravity takes over, the gas fragments into high-density islands ($n \sim 10^5 - 10^8 \rm{cm}^{-3}$) roughly a few hundred AU in size. We refer to these structures as \emph{pseudo-cores}. To identify these regions, we employ a dendrogram algorithm \citep{Rosolowsky2008} on the final simulation snapshot ($t = t_{\rm ff}$), identifying peaks with density $n > 10^4 n_0$. We then tag all tracer particles located within the densest zones of these peaks ($n > n_{\rm peak}^{3/4}$) as "core particles."

This selection procedure isolates $\sim 117,364$ particles ($\sim 5.6\%$ of the total population) that successfully accrete onto prestellar cores. We partition the dataset into two subsets: \emph{Core Particle Dataset} (particles ending in cores) and \emph{Non-Core Particle Dataset} (particles that remain in the diffuse medium). 

To prepare the data for supervised learning, we transform the time-series data into a tabular regression format. The objective of our model $\Phi$ is to map the state vector of a particle at time $t$, $\mathbf{x}_t$, to its future position at time $t + \Delta T$. We construct the training dataset (hereafter referred to as the \emph{Supervised Tabular Dataset}, with its \emph{Core} and \emph{Non-Core} subsets) by pairing the input features at snapshot $i$ with the target coordinates at snapshot $i + \Delta_{\rm frame}$. 
For the primary regression task of forecasting continuous trajectories, the resulting feature vector for each sample is $\mathbf{X} = \{ \text{ID}, \text{Frame}, x, y, z, v_x, v_y, v_z, B_x, B_y, B_z, \log_{10}\rho \}$, and the target vector is $\mathbf{Y} = \{ x_{t+\Delta T}, y_{t+\Delta T}, z_{t+\Delta T} \}$. Hereafter, we will use $\rho_i$ and $Density_i$ to refer to the log-transformed density derived from the simulation, unless stated otherwise. For the supplementary binary classification task (detailed in Section \ref{sec:classification_results}), we further augment this base feature set with locally engineered, rotationally invariant scalars. To explicitly provide the model with flow-field interactions, these additional classification features include the velocity magnitude ($|\mathbf{v}|$), magnetic field magnitude ($|\mathbf{B}|$), their alignment ($\mathbf{v} \cdot \mathbf{B}$ and $\cos \theta$), an Alfv\'enic Mach proxy ($v^2/B^2$), and local kinetic energy ($E_{\rm kin}$).

\begin{figure}[]
\includegraphics[width=\columnwidth]{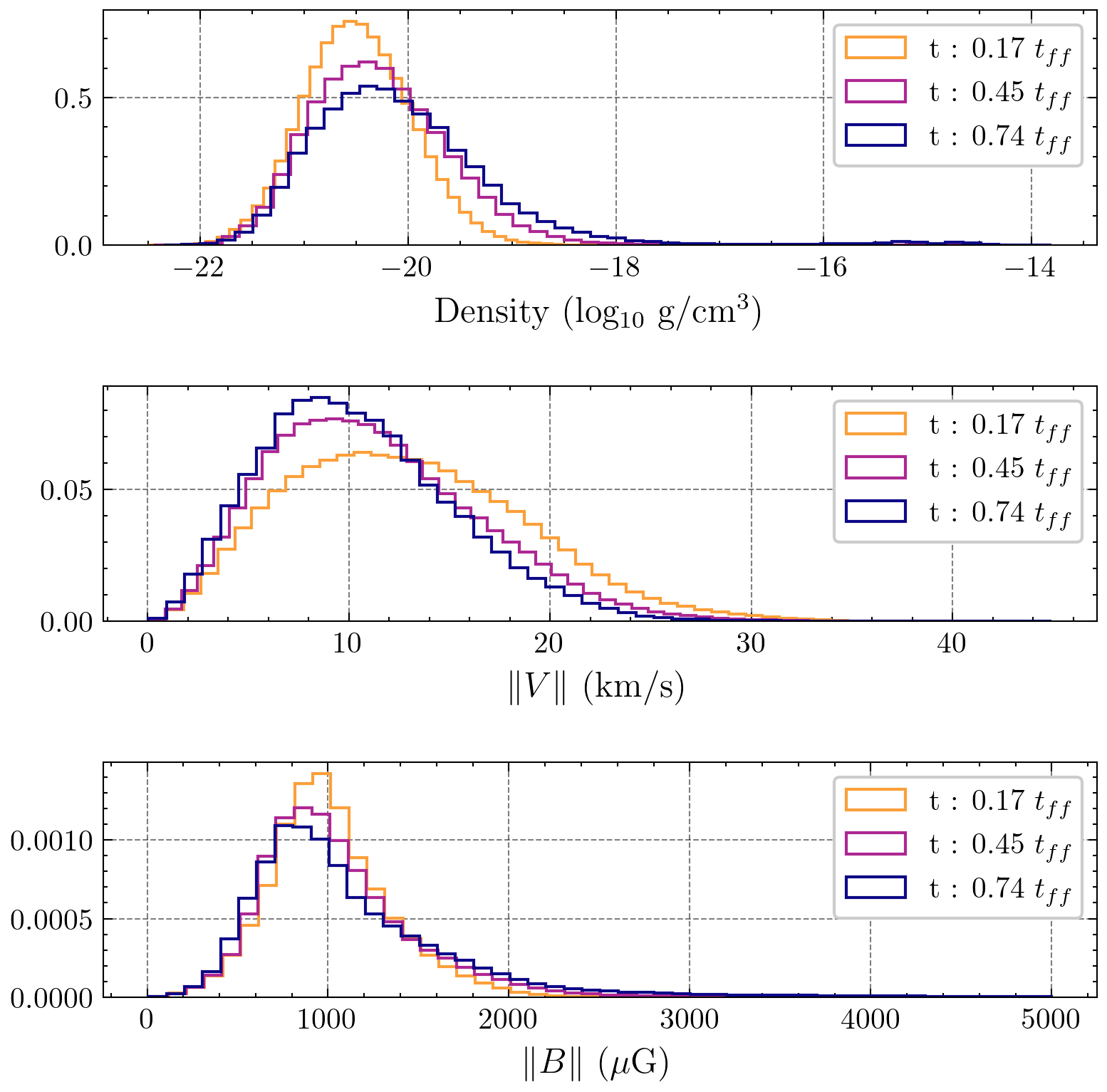}
\caption{Global evolution of the kinematic and magnetic gas properties across the simulation. \textbf{Top:} Probability distribution function (PDF) of log-density at frames 20 (yellow), 55 (purple), and 90 (blue). \textbf{Middle:} PDF of velocity magnitude. \textbf{Bottom:} PDF of the local magnetic field magnitude. As gravitational collapse proceeds, the density develops a prominent power-law tail characteristic of star-forming regions, while the velocity peak sharpens as initial supersonic turbulence decays into coherent infall. The magnetic field strength scales rapidly upward from the initial $13\ \mu\text{G}$ macroscopic background. Due to ideal MHD flux freezing, the field lines are compressed alongside the gas, amplifying the local magnetic field to several milliGauss ($>10^3\ \mu\text{G}$) in the deepest potential wells.
\label{fig:feat_frame}}
\end{figure}
\subsection{Statistical Analysis of Features} \label{sec:stat_var}

Before training, we analyze the statistical properties of the \emph{Core Supervised Tabular Dataset}. Figure \ref{fig:tracks} illustrates the phase-space evolution of three distinct cores. In the early phase ($t \approx 0.5 t_{\rm ff}$), tracers are spatially extended and possess high velocity dispersion. As collapse ensues, the trajectories converge (top row), density rises by orders of magnitude (middle row), and the velocity distribution narrows (bottom row), signifying the transition from turbulent support to gravitational infall.

\begin{figure}[]
\includegraphics[width=\columnwidth]{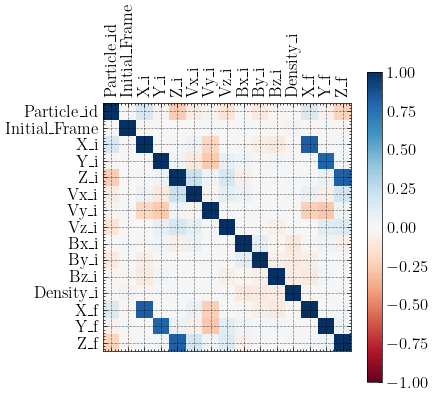}
\includegraphics[width=\columnwidth]{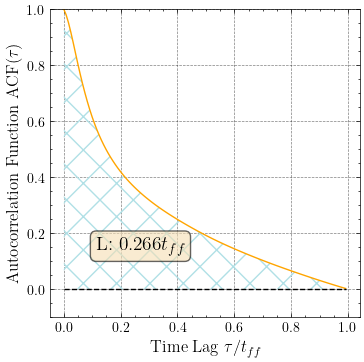}
\caption{\textbf{Top}: Pearson correlation matrix for the training features in \emph{Supervised Tabular Dataset}. \textbf{Bottom}: Temporal autocorrelation function of the density field. The integral of this curve yields the turbulent correlation time, which informs our choice of the prediction horizon $\Delta T$.
\label{fig:corr}}
\end{figure}

Figure \ref{fig:feat_frame} shows the global probability density functions (PDFs) for the primary phase-space scalars density, velocity magnitude, and magnetic field magnitude, at three evolutionary epochs. The density PDF develops a pronounced power-law tail at late times, a hallmark of localized gravitational collapse \citep{Krumholz2005, Collins2012}. Concurrently, the magnetic field distribution shifts toward higher magnitudes (reaching tens to hundreds of $\mu$G), reflecting the amplification of the field via flux freezing as the gas compresses into dense pseudo-cores.
We also compute the Pearson correlation matrix for the primary regression input features (Figure \ref{fig:corr}, top). As expected, the final position is most strongly correlated with the initial position. However, significant non-zero correlations exist between the kinematic, thermodynamic, and newly incorporated magnetic variables ($B_x, B_y, B_z$), which the machine learning model will exploit to improve upon simple ballistic extrapolation.

A critical hyperparameter in this formulation is the prediction horizon, $\Delta T$. A very small $\Delta T$ renders the problem trivial (linear advection), while a very large $\Delta T$ makes the turbulent mapping non-unique and chaotic. We select a lag of $\Delta_{\rm frame} = 30$ snapshots, which corresponds to $\Delta T \approx 0.31$ Myr. This value was chosen to match the integral time scale of the turbulence, ensuring the model predicts evolution over physically significant dynamical times. We define the density fluctuation field $\delta(x,t) = \rho(x,t) - \bar{\rho}$ and calculate its normalized temporal autocorrelation function (ACF):
\begin{equation}
    \mathcal{ACF}(\tau) = \frac{\langle \delta(x,t) \delta(x,t+\tau) \rangle_{x,t}}{\langle \delta(x,t)^2 \rangle_{x,t}}
\end{equation}
The integral time scale $\Delta T$ is obtained by integrating the ACF:
\begin{equation}
    \Delta T = \int_0^{t_{\rm ff}} \mathcal{ACF}(\tau) d\tau
\end{equation}
As shown in Figure \ref{fig:corr} (bottom), this integral converges to approximately $0.25~t_{\rm ff}$, which corresponds to $\sim 0.31$ Myr or 30 simulation snapshots. This confirms that our prediction window matches the coherence time of density structures in the flow.

%% file: 3_Methods.tex
\section{Methods} \label{sec:methods}

\subsection{Problem Formulation} \label{sec:prob_form}

We formulate the prediction of prestellar core formation as a supervised regression problem. Let the state of a tracer particle at time $t$ be represented by a feature vector $\mathbf{x}_t \in \mathbb{R}^n$, containing local physical properties (position, velocity, density, and magnetic field). Our objective is to learn a mapping function $\Phi: \mathbb{R}^n \to \mathbb{R}^3$ that predicts the particle's spatial coordinates at a future time $t + \Delta T$.

Formally, given a dataset $\mathcal{D} = \{(\mathbf{x}_i, \mathbf{y}_i)\}_{i=1}^N$, where $\mathbf{y}_i \in \mathbb{R}^3$ is the true spatial position of the $i$-th particle at the target time, we optimize $\Phi$ by minimizing the total regularized objective function $\mathcal{L}(\Phi)$:
\begin{align}
    \mathcal{L}(\Phi) = \sum_{i=1}^{N} l(\mathbf{y}_i, \Phi(\mathbf{x}_i)) + \Omega(\Phi)
\end{align}
where $l(\mathbf{y}_i, \Phi(\mathbf{x}_i))$ is the pointwise loss function measuring the Euclidean prediction error for sample $i$, and $\Omega(\Phi)$ is a regularization term penalizing model complexity. Unlike simple linear advection schemes, the mapping $\Phi$ must capture the non-linear dynamics of magnetohydrodynamic turbulence, including shocks and gravitational acceleration. 

\textbf{Binary Classification Task:} In addition to continuous trajectory regression, we formulate a supplementary binary classification task to directly predict the ultimate fate of gas parcels. Let $c_i \in \{0, 1\}$ be the binary target label, where $c_i = 1$ if the particle successfully accretes onto a prestellar core at $t \approx t_{\rm ff}$, and $c_i = 0$ otherwise. We represent the classifier model as outputting a raw decision score, in log-odds $f(\mathbf{x}_i) \in (-\infty, +\infty)$ given the feature vector $\mathbf{x}_i \in \mathbb{R}^m$. The predicted probability of core formation, $\hat{p}_i \in [0,1]$, is obtained by applying the logistic sigmoid transformation $\Psi$:
\begin{align}
    \hat{p}_i = \Psi(\mathbf{x}_i) = \frac{1}{1 + e^{-f(\mathbf{x}_i)}}
\end{align}
This task explicitly tests the predictive limits of pointwise kinematic and thermodynamic variables in determining gravitational collapse.

\subsection{Extreme Gradient Boosting (XGBoost)} \label{sec:xgboost}

To approximate $\Phi$ and $\Psi$, we employ Extreme Gradient Boosting (XGBoost) \citep{Chen2016}, an efficient implementation of the Gradient Boosted Decision Tree (GBDT) algorithm. We select a tree-based ensemble over traditional linear regression or simple neural networks for two reasons: (1) Decision trees naturally handle the sharp discontinuities present in supersonic turbulent flows (shocks), and (2) they are invariant to the monotonic scaling of features, requiring less pre-processing of variables like density, which spans many orders of magnitude.

The model approximates the target variable by summing the outputs of $K$ regression trees:
\begin{align}
    \hat{\mathbf{y}}_i = \Phi(\mathbf{x}_i) = \sum_{k=1}^K f_k(\mathbf{x}_i), \quad f_k \in \mathcal{F}
\end{align}
where $\mathcal{F}$ is the space of regression trees. The model is trained additively: at each boosting iteration step $b$, a new tree $f_b$ is added to minimize the residual errors of the prior ensemble. The objective function at step $b$ is:
\begin{align}
    \mathcal{L}^{(b)} = \sum_{i=1}^{N} l(\mathbf{y}_i, \hat{\mathbf{y}}_i^{(b-1)} + f_b(\mathbf{x}_i)) + \Omega(f_b)
\end{align}
The regularization term $\Omega(f)$ penalizes model complexity to ensure physical generalizability:
\begin{align}
    \Omega(f) = \gamma L_{\text{leaves}} + \frac{1}{2} \lambda ||w||^2
\end{align}
where $L_{\text{leaves}}$ is the number of leaves in the tree and $w$ represents the leaf weights. We tune the hyperparameters $\gamma$ (minimum loss reduction) and $\lambda$ (L2 regularization) via grid search.

The choice of the pointwise loss function $l$ depends directly on the predictive task:
\begin{enumerate}
    \item \textbf{Trajectory Regression:} For predicting continuous 3D spatial coordinates using \texttt{XGBRegressor}, $l$ is defined as the half Euclidean squared error ($L_2$ loss):
    \begin{align}
        l(\mathbf{y}_i, \hat{\mathbf{y}}_i) = \frac{1}{2} \|\mathbf{y}_i - \hat{\mathbf{y}}_i\|_2^2
    \end{align}
    \item \textbf{Core Formation Classification:} For predicting the binary collapse fate using \texttt{XGBClassifier}, $l$ is defined as the weighted binary cross-entropy (logarithmic loss):
    \begin{align}
        l(c_i, \hat{p}_i) = - \left[ w_{\text{pos}} \cdot c_i \ln(\hat{p}_i) + (1 - c_i) \ln(1 - \hat{p}_i) \right]
    \end{align}
    where $c_i \in \{0,1\}$ is the true label, $\hat{p}_i \in [0,1]$ is the predicted probability, and $w_{\text{pos}} = 16.87$ is the positive class weighting factor (\texttt{scale\_pos\_weight}) configured to compensate for the extreme volumetric imbalance between diffuse gas and core particles.
\end{enumerate}

\begin{deluxetable*}{lcccccccccc}
\tabletypesize{\scriptsize}
\tablewidth{0pt} 
\tablecaption{Hyperparameter gridsearch results best  and worst models \label{tab:gridsearch}}
\tablehead{
\colhead{Model} & \multicolumn{5}{c}{Hyperparameters}  & \colhead{Mean Absolute Error} & \multicolumn{2}{c}{Time Taken}\\
\colhead{} & \multicolumn{5}{c}{}  & \colhead{$(pc)$} & \multicolumn{2}{c}{$(s)$} & \colhead{} \\
\cline{2-6} \cline{8-9}
\colhead{} & \colhead{\texttt{n\char`_estimators}} & \colhead{\texttt{eta}} & \colhead{\texttt{gamma}} & \colhead{\texttt{subsample}} & \colhead{\texttt{max\char`_depth}} & \colhead{Mean (std)} & \colhead{Mean (std) Fitting} & \colhead{Mean (std) Scoring}} 
\startdata 
{Model A (Optimized)}& 500 & 0.1 & 0.5 & 1.0 & 8 &\textbf{ 0.040926} (0.000101) & 170.4 (20.9) & 2.8 (0.8)\\ 
\hline
{Model B (Intermediate)}& 1000 & 0.01 & 1 & 0.4 & 7 & 0.054045 (0.000069) & 594.7 (109.4) & 20.9 (7.8)\\ 
\hline
{Model C (Baseline)}& 500 & 0.01 & 5 & 0.4 & 7 & 0.070951 (0.000110) & 424.6 (127.2) & 6.5 (2.9)\\ 
\enddata
\end{deluxetable*}
\subsection{Model Evaluation Metrics} \label{sec:model_eval}

We evaluate the regression performance using four statistical metrics. To account for the periodic boundary conditions of our simulation box (where a position of $1.0$ is equivalent to $0.0$), we define the periodic distance function $\mathcal{P}$:
\begin{align}
    \mathcal{P}(a, b) = \min(|a - b|, 1 - |a - b|)
\end{align}

\begin{enumerate}
    \item \textbf{Periodic Mean Absolute Error (P-MAE):}
    The primary metric for optimizing the model.
    \begin{align}
        \text{P-MAE} = \frac{1}{N}\sum_{i=1}^N \mathcal{P}(\mathbf{y}_i, \hat{\mathbf{y}}_i)
    \end{align}

    \item \textbf{Periodic Coefficient of Determination ($R^2_{\rm p}$):}
    Measures the proportion of variance in the particle trajectories captured by the model.
    \begin{align}
        R^2_{\rm p} = 1 - \frac{\sum \mathcal{P}(\mathbf{y}_i, \hat{\mathbf{y}}_i)^2}{\sum \mathcal{P}(\mathbf{y}_i, \bar{\mathbf{y}})^2}
    \end{align}
    where $\bar{y}$ is the mean of the data.

    \item \textbf{Cross-Validation Uncertainty ($\sigma_{\rm CV}$):}
    To estimate the model's reliability on unseen data, we employ repeated k-fold cross-validation (10 folds, 3 repeats) \citep{Stone1974}. The standard deviation of the MAE across these folds, $\sigma_{\rm CV}$, serves as the systematic uncertainty of our model.

    \item \textbf{Bounded Accuracy Fraction:}
    To quantify the reliability of our model beyond global averages, we introduce the \textit{Bounded Accuracy Fraction} ($A_k$). Unlike the Coefficient of Determination ($R^2$), which can be heavily influenced by outliers, $A_k$ measures the percentage of test particles whose prediction error lies within a strict physical tolerance. We define this tolerance based on the intrinsic uncertainty of the model, estimated via the mean absolute error of the cross-validation set ($\sigma_{\rm CV} = \text{MAE}_{\rm CV}$). The Bounded Accuracy for a tolerance factor $k$ is defined as:
    
    \begin{align} \label{eq:bounded_acc}
    A_k = \frac{1}{N} \sum_{i=1}^{N} \mathbb{I} \left( \mathcal{P}(\mathbf{y}_i, \hat{\mathbf{y}}_i) < k \cdot \sigma_{\rm CV} \right)
    \end{align}
    
    where $\mathbb{I}$ is the indicator function. In this work, we report results for $k=3$. This corresponds to a $3\sigma$ confidence interval, a standard threshold in physical sciences for robust statistical significance. A high value of $A_3$ indicates that the model's errors are well-behaved and that "catastrophic" failures (predictions deviating by $>3\times$ the typical error) are rare.

\end{enumerate}
\textbf{Classification Metrics:}\\
To evaluate the binary classification model, we utilize standard metrics suited for highly imbalanced datasets. We define True Positives (TP) as correctly identified core particles, False Positives (FP) as non-core gas incorrectly flagged for collapse, and False Negatives (FN) as missed core particles.
\begin{itemize}
    \item \textbf{Recall (True Positive Rate):} $\text{TP} / (\text{TP} + \text{FN})$. Measures the model's sensitivity in identifying gravitationally susceptible gas.
    \item \textbf{Precision:} $\text{TP} / (\text{TP} + \text{FP})$. Measures the reliability of the model's collapse predictions, directly exposing the false-positive rate.
    \item \textbf{Area Under the Curve (AUC):} We compute the AUC for both the Receiver Operating Characteristic (ROC) and Precision-Recall (PR) curves to quantify the model's diagnostic ability across all probability thresholds.
\end{itemize}

\subsection{Model Interpretability via SHAP} \label{sec:shap_methods}

To unpack the decision-making logic of our trained XGBoost models and ensure that predictions are driven by physically meaningful signals rather than spatial artifacts, we employ SHapley Additive exPlanations (SHAP) \citep{Lundberg2017}. Grounded in cooperative game theory \citep{Shapley2016}, SHAP provides a mathematically rigorous framework for feature attribution by calculating the marginal contribution of each input variable across all possible feature subsets.

For an input feature vector $\mathbf{x}$, the SHAP attribution value $\phi_j(\mathbf{x})$ assigned to feature $j \in F$ is defined as:
\begin{align}
    \phi_j(\mathbf{x}) = \sum_{S \subseteq F \setminus \{j\}} \frac{|S|!(|F| - |S| - 1)!}{|F|!} \left[ f_x(S \cup \{j\}) - f_x(S) \right]
\end{align}
where $F$ is the complete set of input features, $S$ is a subset of features excluding feature $j$, and $f_x(S)$ is the conditional expectation of the model log-odds output given feature subset $S$.

SHAP values satisfy local accuracy (additivity), meaning the sum of all feature attributions equals the difference between the particle's specific log-odds score $f(\mathbf{x})$ and the dataset-wide base rate expectation $E[f(\mathbf{X})]$:
\begin{align}
    f(\mathbf{x}) = E[f(\mathbf{X})] + \sum_{j=1}^{|F|} \phi_j(\mathbf{x})
\end{align}
Here, $E[f(\mathbf{X})]$ represents the prior log-odds expectation across the background cloud population (which is strongly negative owing to the severe rarity of collapsing gas, $\sim 5.6\%$). Positive SHAP values ($\phi_j > 0$) indicate that a feature pushes the gas parcel toward gravitational collapse ($p > 0.5$), whereas negative SHAP values ($\phi_j < 0$) indicate that a feature reinforces diffuse non-core behavior.

Because exact Shapley calculations are computationally prohibitive for large datasets, we utilize \texttt{TreeExplainer}\footnote{https://github.com/shap/shap} \citep{Lundberg2020}, a high-performance algorithm optimized for decision tree ensembles. \texttt{TreeExplainer} evaluates exact SHAP values in polynomial time ($O(TLD^2)$, where $T$ is the number of trees, $L$ is maximum leaves, and $D$ is maximum depth) by tracking conditional expectations down tree decision paths. 

We evaluate SHAP attributions across the out-of-fold stratified test set to derive three complementary insights:
\begin{enumerate}
    \item \textbf{Global Feature Importance:} Measured by the mean absolute SHAP value across all $N$ test particles, $\frac{1}{N}\sum_{i=1}^N \vert{}\phi_j(\mathbf{x}_i)\vert{}$, establishing the hierarchical ranking of physical variables governing core identification.
    \item \textbf{Non-linear Dependence \& Interactions:} Scatter dependence plots mapping feature values $\mathbf{x}_{j}$ against attributions $\phi_j$, color-coded by interacting secondary variables to reveal physical phase transitions (e.g., density collapse thresholds and magnetic support thresholds).
    \item \textbf{Local Particle Attribution:} Individual waterfall decompositions showing how local thermodynamic and magnetic properties cumulatively shift single tracer particles away from the background expectation $E[f(\mathbf{X})]$ into core classification.
\end{enumerate}

%% file: 4_Results.tex
\section{Results} \label{sec:results}

\subsection{Hyperparameter Optimization} \label{sec:gridsearch}

We construct tree-boosted regression models using the \textbf{XGBoost} framework. To ensure our model generalizes well to unseen turbulent flow configurations and avoids overfitting to specific simulation snapshots, we perform a rigorous hyperparameter optimization. We utilize repeated k-fold cross-validation (3 repeats, 10 folds) on a training subset of $2.3 \times 10^6$ particle trajectories. This ensures that the validation fold is strictly disjoint from the training data, mimicking the evaluation of an unknown physical state.

The hyperparameter search space was selected to balance model complexity with computational efficiency, given the high-dimensional nature of turbulent flows:
\begin{itemize}
    \item \texttt{n\char`_estimators} $\in [500, 1500]$: The number of boosted trees. We test lower bounds to avoid underfitting complex shocks and upper bounds to limit computational overhead.
    \item \texttt{eta} (Learning Rate) $\in [0.01, 0.3]$: Controls the step size of the gradient descent. Lower values ($0.01$) generally yield more robust models but require more trees.
    \item \texttt{gamma} $\in [0.5, 5]$: The minimum loss reduction required to make a further partition on a leaf node. Higher values act as regularization, preventing the model from fitting numerical noise in the low-density regions.
    \item \texttt{subsample} $\in [0.4, 1.0]$: The fraction of observations to be randomly sampled for each tree. Subsampling introduces randomness that improves robustness against local outliers in the simulation.
    \item \texttt{max\char`_depth} $\in [6, 8]$: Constrains the depth of individual trees. We cap this at 8 to prevent the model from memorizing specific particle trajectories.
\end{itemize}

The results of the grid search are summarized in Table \ref{tab:gridsearch}. We designate the highest performing configuration as \textbf{Model A (Optimized)}, which utilizes a high subsample rate (1.0) and moderate depth (8), achieving the lowest Mean Absolute Error (MAE). Models with aggressive regularization (high \texttt{gamma}, low \texttt{eta}) generally underperformed (\textbf{Model C}), suggesting that the complexity of turbulent core collapse requires a model with sufficient capacity to capture fine-grained phase-space correlations.
\begin{figure}[]
\includegraphics[width=\columnwidth]{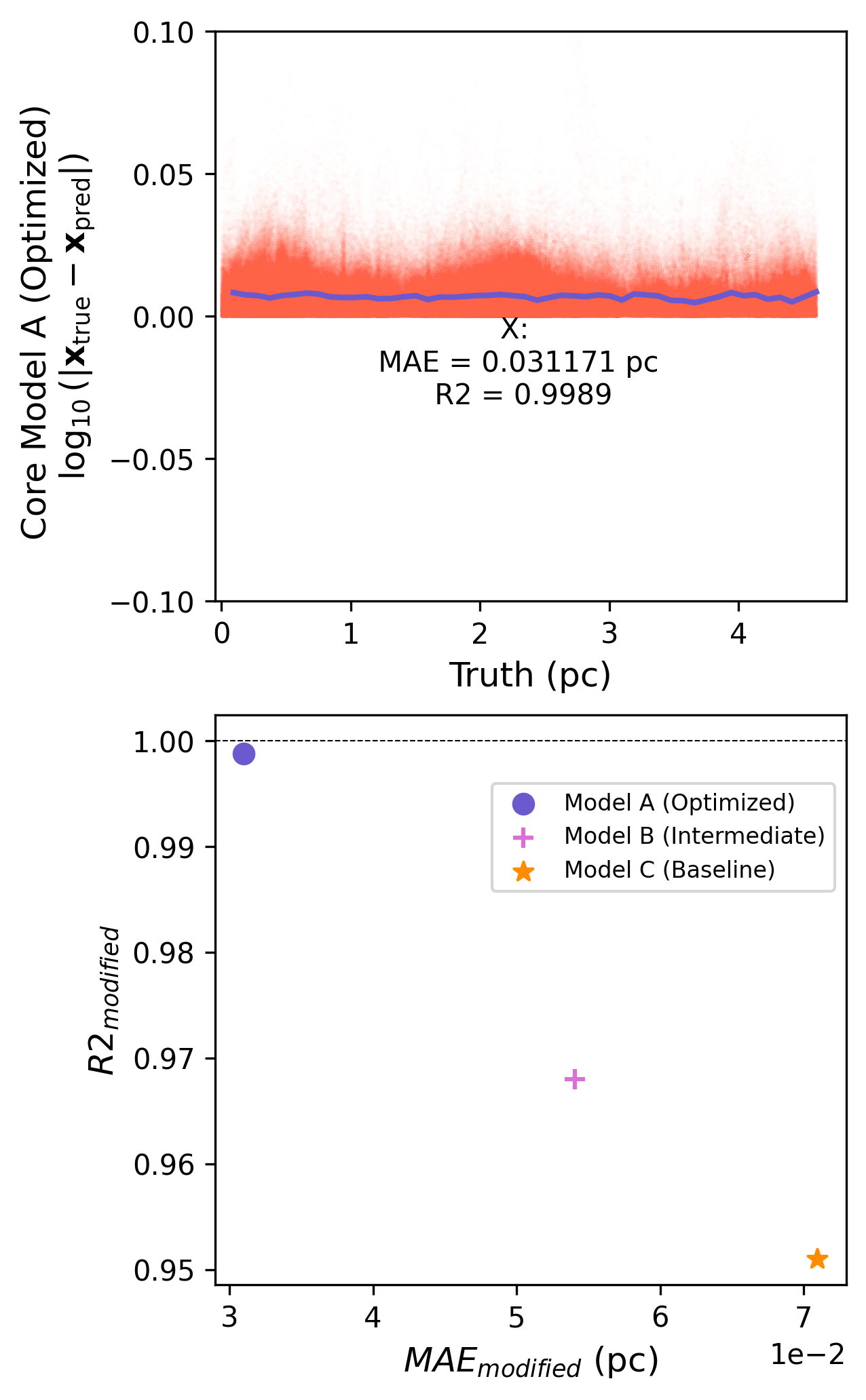}
\caption{Performance comparison of the Optimized (Model A), Intermediate (Model B), and Baseline (Model C) architectures. \textbf{Top:} Prediction-Truth for X position of Core particles across the test set. \textbf{Bottom:} The coefficient of determination ($R^2$) versus MAE for the different hyperparameter configurations. While the Baseline Model C suffers from high variance (scatter), the Optimized Model A consistently achieves sub-resolution accuracy with a tighter error distribution, demonstrating the necessity of deeper trees to capture turbulent intermittency.
\label{fig:gen_pred}}
\end{figure}

\subsection{Global Model Performance \& Reliability} \label{sec:em_1}

Having identified the optimal hyperparameter configuration (Model A) and two reference points (Intermediate Model B and Baseline Model C), we evaluate their global predictive capabilities. A critical question is not just minimizing the mean error, but ensuring the model is reliable across different physical regimes (e.g., high-density cores vs. diffuse gas).

Figure \ref{fig:gen_pred} compares the performance of the three model tiers. \textbf{Model A (Optimized)} achieves a global MAE of $0.0312$ pc ($R^2=0.9988$). In contrast, the \textbf{Baseline Model C} (characterized by high regularization and shallow trees) shows a $40\%$ higher error rate ($MAE \approx 0.069$ pc). 

To further quantify the predictive value of our chosen features, we evaluated the impact of incorporating local magnetic field components ($B_x, B_y, B_z$) alongside the kinematics, and contrasted this against an ablation study using a reduced "Position-Only" model restricted strictly to spatial coordinates ($x, y, z$). Including the localized magnetic field vectors yielded minor but consistent improvements to the overall continuous regression accuracy. While short-term trajectory advection is heavily dominated by bulk kinematics, the magnetic features assist in fine-tuning trajectories in highly magnetized environments. Conversely, the "Position-Only" model degraded significantly in high-density core regimes. Without local velocity and density information, the model failed to anticipate the non-linear acceleration driven by gravitational collapse, confirming that the full phase-space vector is essential for accurately forecasting star formation dynamics.

\begin{figure}[]
\includegraphics[width=\columnwidth]{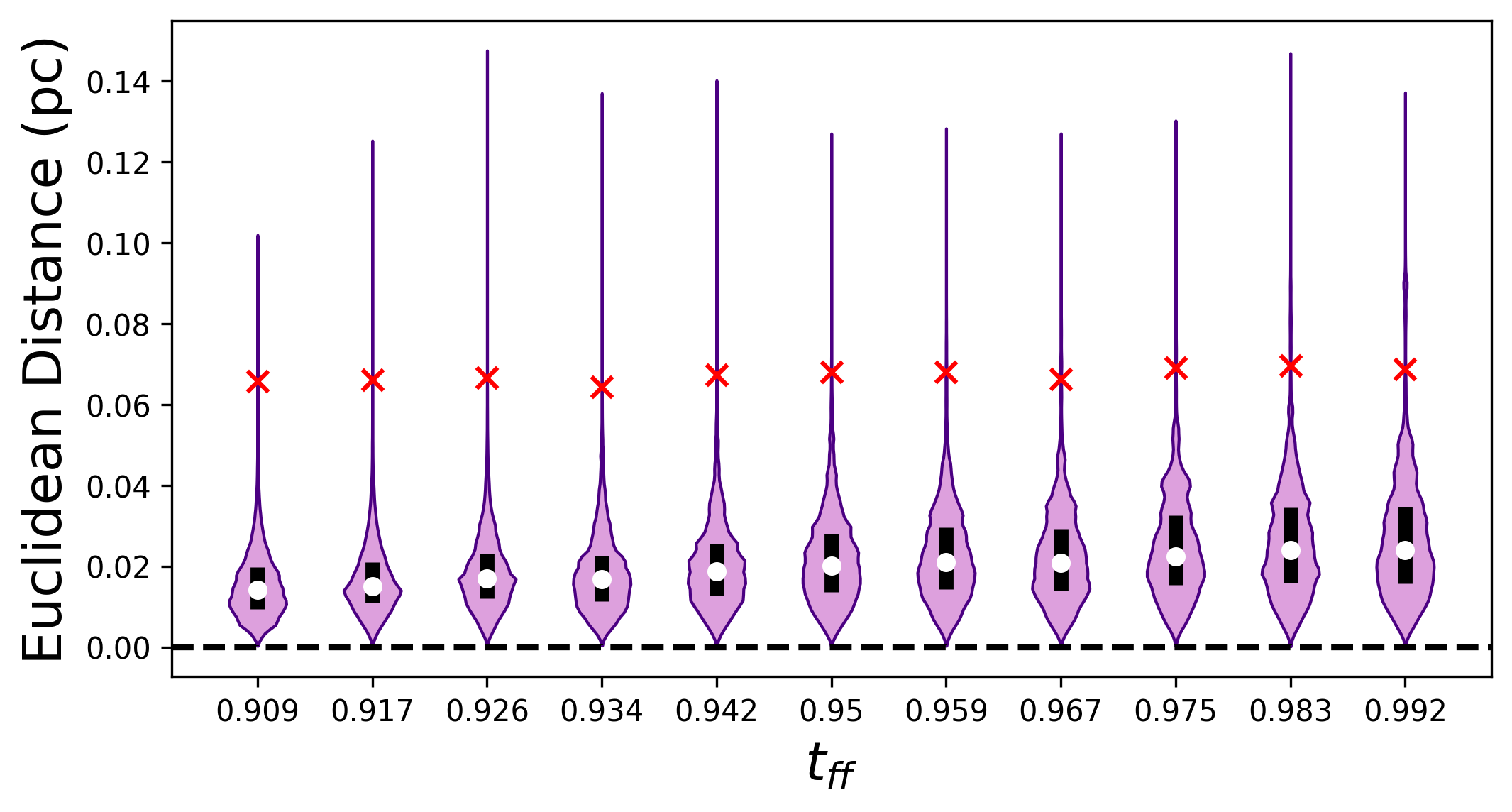}
\includegraphics[width=\columnwidth]{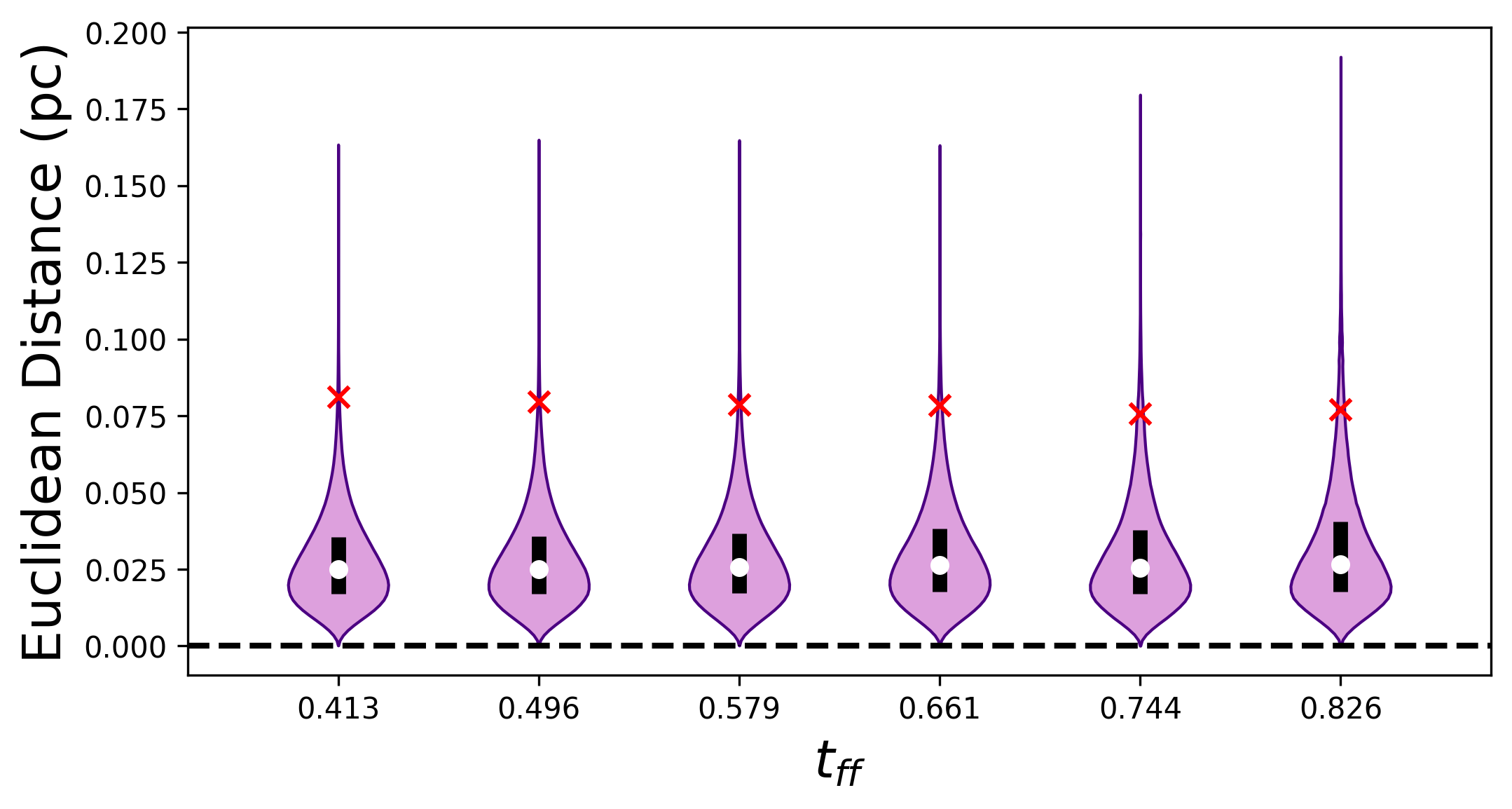}
\caption{Temporal generalization and robustness testing. \textbf{Top:} Distribution of framewise prediction errors (Euclidean distance) for the test set within the training simulation ($u501$). The XGBoost predictions (Purple) consistently maintain lower errors than the No-Motion Baseline (Red Crosses), which assumes static particles over the prediction interval. \textbf{Bottom:} Generalization to an independent simulation run ($u502$) initialized with distinct turbulence seeds. Although the error marginally increases for the unseen simulation, the model retains predictive power ($50-60\%$ improvement over baseline), demonstrating that it has learned physical advection rules rather than memorizing the training volume.
\label{fig:frame_pred}}
\end{figure}

\subsection{Temporal Generalization and Cross-Simulation Robustness} \label{sec:frame_pred}

While global error metrics provide a baseline for model performance, the true test of a predictive physical model is its ability to forecast future states and generalize to unseen initial conditions. To evaluate this, we perform a "future frame" extrapolation test. The model, trained strictly on the early evolutionary phase ($t \in [0.167, 0.67] t_{ff}$) of our fiducial simulation (\textit{u501}), is tasked with predicting the phase-space coordinates of particles in a future time interval ($t \in [0.925, 1] t_{ff}$).

Figure \ref{fig:frame_pred} is the central result of this temporal validation. The top panel displays the distribution of the Euclidean distance error ($| \mathbf{x}_{\rm true} - \mathbf{x}_{\rm pred} |$) for the XGBoost predictions (purple) compared against a "No-Motion Baseline" (red crosses). The No-Motion baseline assumes particles remain static over the prediction horizon ($\Delta T$), effectively measuring the inherent bulk flow of the gas. Beating this baseline is the fundamental criterion for proving that the machine learning algorithm has learned the physical kinematics of the system rather than merely applying a persistence forecast. As shown, the model significantly outperforms the baseline across all future frames. While the prediction error naturally grows as the simulation evolves further from the training window which is an unavoidable consequence of the positive Lyapunov exponents in chaotic turbulent flows, the XGBoost predictions remain tightly bounded and physically realistic.

\subsubsection{Generalization to Varying Magnetic Field Strengths}
To rigorously test the generalizability of our learned phase-space correlations, we deploy the model on a completely independent simulation run, \textit{u502}. Crucially, \textit{u502} is not merely a different random realization of the turbulent velocity field; it features a different initial magnetic field strength. While the training simulation (\textit{u501}) is strongly magnetized with a 
plasma beta of $\beta = 0.2$, the validation simulation (\textit{u502}) is moderately magnetized with $\beta = 2.0$. Aside from the altered magnetic field, \textit{u502} was initialized with identical global parameters ($\mathcal{M} = 9$, $\alpha_{\rm vir} = 1$) but a distinct random seed, resulting in completely unique morphological structures, such as different filament locations and core formation sites.

The bottom panel of Figure \ref{fig:frame_pred} illustrates the framewise prediction error when the \textit{u501}-trained model is applied directly to \textit{u502}. Despite the altered magnetic pressure support and different turbulent morphology, the model successfully generalizes. While the median error distribution shifts marginally higher compared to the self-consistent test, the model maintains a $50-60\%$ improvement over the No-Motion baseline. It achieves a robust Mean Absolute Error of $\approx 0.012$ pc and maintains a Bounded Accuracy ($A_3$) of over $85\%$. This result strongly implies that the core kinematic signatures of gravitational instability like local density enhancements coupled with convergent velocity flows, are relatively invariant across different global magnetic field strengths, allowing our kinematic and local magnetic feature set to maintain predictive power. This confirms that the XGBoost ensemble has learned the generalized physical rules of turbulent advection and collapse, rather than simply memorizing the specific spatial geometry of the training box.
\begin{figure*}[]
\includegraphics[width=2\columnwidth]{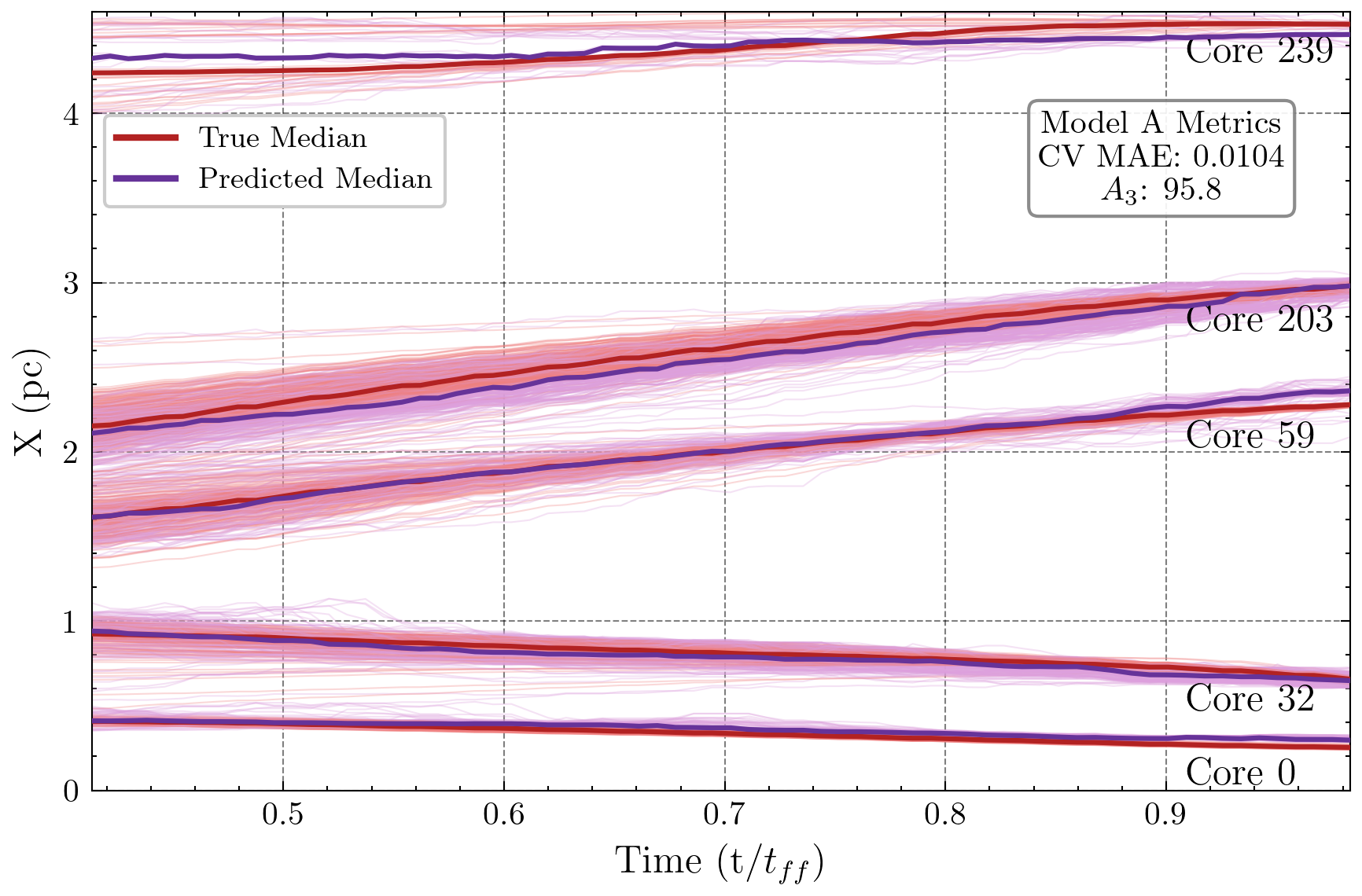}
\caption{Trajectory reconstruction for five representative prestellar cores. The solid lines represent the median path of the constituent gas parcels (Red: Ground Truth, Purple: Model Prediction), while the faint lines show a subset of individual particle traces. The model successfully recovers the convergent flow and bulk motion of the cores over a 0.45 Myr horizon. Model A achieves a Bounded Accuracy of $A_3 \approx 96\%$, indicating that the vast majority of predicted trajectories fall within the $3\sigma$ confidence interval of the cross-validation error.
\label{fig:corewise_pred}}
\end{figure*}
\subsection{Trajectory Reconstruction (Corewise Prediction)} \label{sec:core_pred}

The ultimate test for our application is the ability to reconstruct the formation history of prestellar cores. We select identifiable cores from the final simulation snapshot and recursively predict their trajectories backward and forward in time using the test set partition. To quantify the model's reliability, we calculate the Bounded Accuracy Fraction ($A_k$) as defined in Eq. \ref{eq:bounded_acc}. Using a strict $3\sigma$ equivalent threshold ($k=3$), we find that \textbf{Model A achieves an accuracy of $A_3 \approx 96\%$}. This indicates that for more than 9 out of 10 particles, the predicted trajectory lies within three times the typical model uncertainty. In contrast, the baseline Model C drops to $A_3 \approx 65\%$, indicating that nearly a third of its predictions are significant outliers. This confirms that the deeper trees in Model A are essential for capturing the intermittent, non-linear events characteristic of turbulent collapse.

Figure \ref{fig:corewise_pred} visualizes the predicted $x$-position tracks (purple) versus the ground-truth simulation tracks (red) for a random sample of cores. The model successfully recovers the complex, non-linear bulk motion of the cores. Crucially, it captures the convergent flow of individual tracers as they collapse into the potential well. The strong spatial correspondence between the predicted and true mean paths confirms that the algorithm can be used as a reliable "Lagrangian tracker" to identify the provenance of star-forming gas in large-scale simulations where high-frequency output might be unavailable.

\subsection{Identifying Core Formation: The Physical Limits of Point-Wise Data} \label{sec:classification_results}

\begin{figure*}[]
\includegraphics[width=2\columnwidth]{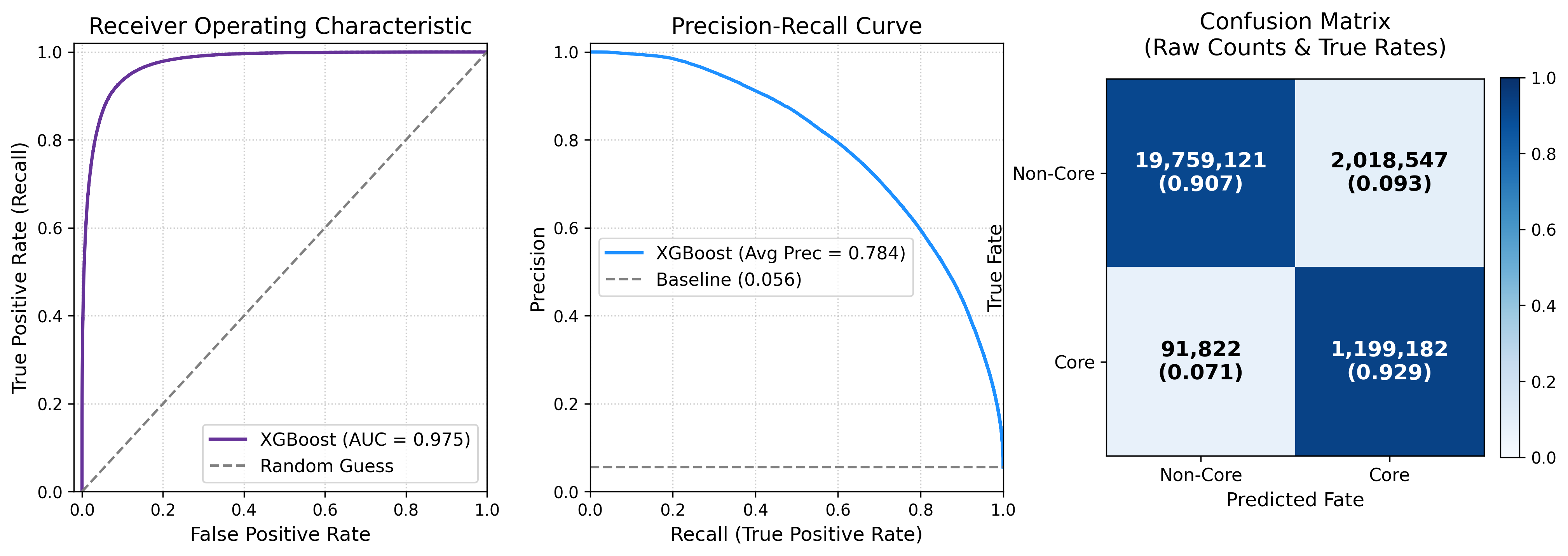}
\caption{Performance of the binary classification model for predicting prestellar core formation across the accumulated test set ($23,068,672$ total particle evaluations spanning 11 temporal snapshots). \textbf{Left:} The Receiver Operating Characteristic (ROC) curve, demonstrating excellent global ranking ability with an AUC of 0.975. \textbf{Middle:} The Precision-Recall (PR) curve, showing an Average Precision of 0.784, a strong result given the severe volumetric class imbalance (only $\sim 5.6\%$ of the cloud undergoes collapse). \textbf{Right:} The dual-annotated confusion matrix evaluated at a binary probability threshold of $p > 0.5$, displaying both raw particle counts and row-normalized recovery rates (in parentheses). While the percentages on the diagonal highlight a high recovery rate of actual star-forming gas (93\% Recall), the raw counts reveal the true physical volume of the false positives. Because the background non-core gas is overwhelmingly abundant, the seemingly small 9.3\% error rate translates into millions of false alarms, ultimately dropping the model's absolute Precision to 37\%.
\label{fig:classification}}
\end{figure*}

To evaluate our supplementary binary classification task of predicting whether a given gas parcel will successfully accrete into a prestellar core, we utilized the augmented feature set containing explicitly engineered kinematic and local magnetic scalars ($|\mathbf{v}|$, $|\mathbf{B}|$, $\mathbf{v} \cdot \mathbf{B}$, $v^2/B^2$, and $E_{\rm kin}$). Because only 5.6\% of the particles form cores, we applied a scale positive weight during training to penalize missed collapse events.

Evaluating the model across all continuous probability thresholds yields exceptional global metrics, as shown in Figure \ref{fig:classification}. The model achieves a Receiver Operating Characteristic (ROC) Area Under the Curve (AUC) of 0.975, and an Average Precision of 0.784. This indicates that the engineered pointwise features possess significant distinguishing power, successfully ranking the gravitational susceptibility of the gas.

However, when evaluating the absolute class predictions at the default probability threshold ($p > 0.5$), the model achieves a high recall of 93\% but a limited precision of 37\% (see Figure \ref{fig:classification}, right panel). While the model successfully flags the vast majority of converging flows, it systematically over-predicts core formation in boundary cases, resulting in a high false-positive rate. 

From a physical standpoint, this threshold behavior is highly illuminating. Gravitational collapse is governed not merely by local state magnitudes, but by differential spatial forces and multi-scale force balance across surrounding gas parcels (e.g., thermal and magnetic pressure gradients, Lorentz forces, and local gravitational potentials). Because the XGBoost model evaluates every Lagrangian tracer particle in pointwise isolation, it lacks the environmental spatial context required to resolve continuous spatial gradients and local force distributions. The tree ensemble probabilistically recognizes dense, magnetized converging gas, but mathematically cannot determine whether neighboring parcels provide sufficient pressure or magnetic support to halt the collapse. Consequently, when forced to make a binary decision, the algorithm conservatively flags susceptible gas as collapsing. This demonstrates that while pointwise features provide excellent probabilistic rankings, achieving high absolute precision in turbulent MHD regimes requires spatially aware machine learning architectures (such as Graph Neural Networks or Convolutional Networks) capable of processing non-local neighborhood context and continuous spatial derivatives.

\subsection{Physical Interpretation via SHAP Analysis} \label{sec:shap_analysis}
\begin{figure*}[t]
\centering
\includegraphics[width=2\columnwidth]{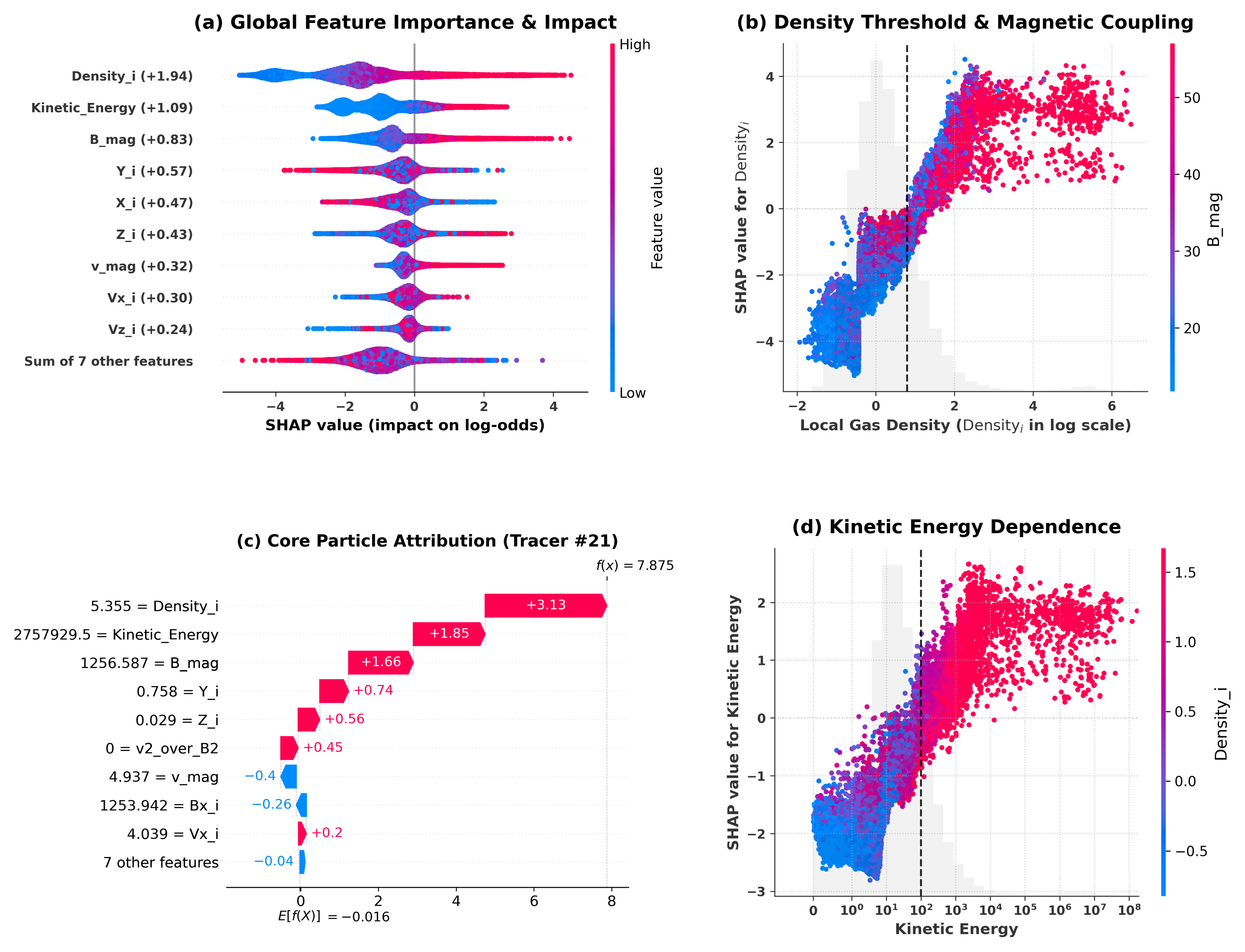}
\caption{SHapley Additive exPlanations (SHAP) framework interpreting the physical drivers governing XGBoost prestellar core predictions. \textbf{(a)} Annotated beeswarm plot of global feature importance, ranking feature impacts on prediction log-odds alongside mean absolute SHAP values in parentheses. Local gas density ($\text{Density}_i$, $+1.94$), kinetic energy ($\text{Kinetic\_Energy}$, $+1.09$), and magnetic field strength ($B_{\rm mag}$, $+0.83$) dominate classification decision rules. \textbf{(b)} Scatter dependence plot of $\text{Density}_i$ color-coded by $B_{\rm mag}$, revealing a sharp non-linear collapse transition at $\text{Density}_i \approx 0.8$ (in logarithmic units relative to mean cloud density; dashed black line). Beyond this threshold ($\text{Density}_i > 2.0$), strong magnetic fields ($B_{\rm mag} \gtrsim 50$) act as a positive multiplier for core identification. \textbf{(c)} Waterfall attribution plot for a representative true-positive core tracer particle (Tracer \#21), illustrating how feature attributions sum from the cloud baseline $E[f(\mathbf{X})] = -0.016$ up to a net log-odds decision score of $f(\mathbf{x}) = 7.875$ ($p > 0.999$). Extreme density (+3.13), kinetic energy (+1.85), and magnetic field magnitude (+1.66) cumulatively push the particle into the core classification regime. \textbf{(d)} Scatter dependence plot of $\text{Kinetic\_Energy}$ color-coded by $\text{Density}_i$ (with background density histogram), demonstrating a key energy threshold at $\text{Kinetic\_Energy} \approx 10^2$ (dashed black line). Below $10^2$, low kinetic energy and ambient densities ($\text{Density}_i < 0.5$) produce negative attributions ($\text{SHAP} \le 0$), whereas above $10^2$, high kinetic energy coupled with dense gas ($\text{Density}_i > 1.0$) strongly drives positive core attribution ($\text{SHAP} \approx +2.0$).}
\label{fig:shap_analysis}
\end{figure*}

To determine the physical mechanisms driving the classifier's decisions and ensure the model is leveraging physically meaningful signals rather than spatial artifacts, we performed a SHapley Additive exPlanations (SHAP) analysis using \texttt{TreeExplainer}. SHAP calculates the marginal contribution of each feature to the model's predicted log-odds across a stratified sample of test particles.

The global feature importance (measured by mean absolute SHAP value; Figure~\ref{fig:shap_analysis}a) reveals that local gas density ($\text{Density}_i$, $+1.94$) is the primary deterministic feature for core identification, followed by total kinetic energy ($\text{Kinetic\_Energy}$, $+1.09$) and magnetic field magnitude ($B_{\rm mag}$, $+0.83$). Spatial position coordinates ($X_i, Y_i, Z_i$) and directional vector components contribute secondary effects, confirming that the tree ensemble relies predominantly on scalar thermodynamic and magnetic state variables rather than spatial localization within the simulation domain.

Scatter dependence analyses highlight key non-linear physical thresholds learned by the model (Figure~\ref{fig:shap_analysis}b, d):
\begin{enumerate}
    \item \textbf{Density Thresholding}: The SHAP value for $\text{Density}_i$ displays a sharp transition at $\text{rho} \approx 0.8$ (in logarithmic units, equivalent to $\approx 6.3 \times$ background density). Below this value, local density negatively impacts the core classification probability (SHAP values down to $-5.0$), whereas above this threshold, density contributions turn strongly positive, peaking near $+4.0$ (Figure~\ref{fig:shap_analysis}b).
    \item \textbf{Magnetic Amplification}: Interaction color-coding shows that in the high-density regime ($\text{Density}_i > 2.0$), high magnetic field magnitudes ($B_{\rm mag} \gtrsim 50$) consistently yield the highest positive SHAP values. This indicates the model learned that strong magnetic fields within dense gas are a strong indicator of bound prestellar structures.
    \item \textbf{Kinetic Energy Thresholding}: The attribution for kinetic energy exhibits a clear physical transition at $\text{Kinetic\_Energy} \approx 10^2$ (Figure~\ref{fig:shap_analysis}d). Below $10^2$, low kinetic energy coincides with low ambient gas density ($\text{Density}_i < 0.5$) to yield negative SHAP values ($\text{SHAP} \le 0$). Above $10^2$, elevated kinetic energy coupled with high density ($\text{Density}_i > 1.0$) drives positive attribution up to $+2.5$, reflecting active infalling gas motions associated with gravitational collapse.
\end{enumerate}

Waterfall attributions for individual true-positive core predictions demonstrate how local physical variables combine additively on a per-particle basis (Figure~\ref{fig:shap_analysis}c; see Section~\ref{sec:shap_methods} for mathematical details). Model predictions are evaluated in log-odds space, where $E[f(\mathbf{X})] = -0.016$ represents the expected baseline score across the cloud background, and positive SHAP attributions ($\phi_j > 0$) actively drive the particle toward core formation. For a representative collapsing tracer particle (Tracer \#21), the model begins at this diffuse baseline ($E[f(\mathbf{X})] = -0.016$). Extreme gas density ($\text{Density}_i = 5.355$) provides the primary positive push (+3.13), which is strongly reinforced by elevated kinetic energy ($\text{Kinetic\_Energy} = 2.76 \times 10^6$, $+1.85$) and magnetic field strength ($B_{\rm mag} = 1256.6$, $+1.66$). Summing all feature attributions yields a net model decision score of $f(\mathbf{x}) = 7.875$, which maps via Equation~(2) to a near-certain core probability of $p = \sigma(7.875) > 0.999$.

These results demonstrate that the XGBoost classifier successfully extracts fundamental MHD collapse dynamics directly from pointwise data, providing physically interpretable decision boundaries alongside strong predictive performance.

%% file: 5_Discussion.tex
\section{Discussion} \label{sec:discussion}

In this work, we have demonstrated that the formation of prestellar cores in turbulent molecular clouds can be accurately forecasted using supervised machine learning. By transforming the Lagrangian history of tracer particles into a tabular regression problem, we trained an Extreme Gradient Boosting (XGBoost) model to predict 3D particle trajectories over physically significant timescales ($\sim 0.25~t_{\rm ff}$).

Our results indicate that the local point-wise information (position, velocity, density, and local magnetic field) contained within a single simulation snapshot holds significant predictive power regarding the future evolution of the gas. Through both continuous trajectory forecasting and binary classification, the model successfully distinguished between transient density fluctuations and gravitationally bound collapsing gas, achieving a high coefficient of determination ($R^2 > 0.99$) for continuous advection and demonstrating exceptional probabilistic ranking capabilities for identifying star-forming regions.

\subsection{Implications for Subgrid Modeling} \label{sec:subgrid}

A primary motivation for this work is the potential to accelerate multi-scale astrophysical simulations. Current cosmological and galaxy-scale simulations (e.g., IllustrisTNG, EAGLE, FIRE) lack the spatial resolution required to follow the gravoturbulent collapse of individual molecular cores ($< 0.1$ pc). Instead, they rely on heuristic subgrid recipes with ad hoc probabilistic criteria based on local density thresholds to spawn star particles. Our XGBoost framework offers a data-driven alternative to these varying prescriptions, using the high-resolution MHD physics learned from simulation data to construct a fast, physical surrogate model.

The choice of a supervised tabular formulation over unsupervised, weakly supervised, or physics-informed frameworks is driven by the specific physical objective of this work. While unsupervised or self-supervised learning (SSL) frameworks excel at discovering latent structural clustering, prestellar core formation represents a rare, non-linear physical outcome affecting only a small fraction ($\lesssim 1\%$) of turbulent gas. Without explicit ground-truth labels to supervise the mapping from initial states to final accretion fates, unsupervised methods tend to cluster gas parcels by dominant ambient kinematics rather than isolating the delicate threshold conditions governing gravitational collapse. Similarly, weakly supervised learning relies on coarse, regional, or aggregate labels; because high-fidelity MHD simulations provide exact Lagrangian particle trajectories, adopting weak supervision would unnecessarily discard fine-grained kinetic and magnetic trajectory data.

Crucially, the requirement for future labels applies exclusively to the offline training phase. During inference within target applications, no future labels are required. Because the inference time of a trained decision tree ensemble is negligible compared to the computational cost of solving the Poisson and MHD fluid equations, this pre-trained model can be implemented "on-the-fly" within lower-resolution simulations, transforming instantaneous fluid states at time $t$ into probabilistic accretion fates over a short-term horizon ($\sim 0.25~t_{\rm ff}$).

In practical subgrid applications, this surrogate model can be deployed via a three-step workflow:
\begin{enumerate}
    \item \textbf{State Extraction:} At a given simulation timestep $t$, the code extracts the instantaneous fluid state vector, comprising local density $\rho$, velocity $\mathbf{v}$, and magnetic field strength $\mathbf{B}$, for a gas parcel.
    \item \textbf{Surrogate Evaluation:} The state vector is passed directly to the pre-trained XGBoost model, which outputs the collapse probability $\hat{y}$ in fractions of a millisecond, bypassing the need to solve computationally expensive sub-grid MHD equations.
    \item \textbf{Subgrid Action:} The simulation code utilizes $\hat{y}$ to govern subgrid physical prescriptions, such as seeding sink particles, determining local Star Formation Rates (SFR), or injecting localized momentum feedback.
\end{enumerate}

In codes employing active subgrid tracer particles (e.g., \textsc{Arepo}, \textsc{Gizmo}, \textsc{Enzo}), the model directly ingests interpolated tracer attributes. In standard Eulerian grid codes, the cell-centered volume-averaged quantities $(\bar{\rho}, \mathbf{\bar{v}}, \mathbf{\bar{B}})$ serve as the surrogate inputs, matching the point-wise interpolation scheme used to construct the training dataset. 

While the model presented here is trained on a single value of Mach number and forcing parameter and will not generalize to a full-physics galaxy simulation, this work serves as a proof-of-concept that decision-tree-based surrogates can predict the outcome of gravoturbulent collapse. In order to build a subgrid model applicable to general galaxy formation, a large suite of collapse simulations covering varied turbulent and magnetic regimes will be required; this will be the focus of the second paper in this series. The advantage of this technique is that it establishes a subgrid foundation that is physically grounded and far more sophisticated than traditional density-threshold models.

\subsection{Strengths and Limitations} \label{sec:P_n_C}

\subsubsection{Strengths}
The primary advantage of this approach is its ability to capture \textbf{non-linear phase-space correlations} without explicit programming. Unlike analytical models that often assume spherical symmetry (e.g., Bonnor-Ebert spheres) or simple Jeans instability thresholds, XGBoost learns the complex, anisotropic geometry of turbulent accretion flows.

Additionally, the use of \textbf{Lagrangian tracer particles} is a significant improvement over Eulerian analysis for this specific task. Grid-based analysis smears out the history of the gas due to advection errors and cell mixing. By defining the problem in the Lagrangian frame, we preserve the identity of the gas parcels, allowing the ML model to "see" the coherent assembly of mass over time. This confirms that tabular machine learning methods can rival complex 3D Convolutional Neural Networks (CNNs) for specific tasks, often with significantly lower computational overhead and training data requirements.

Furthermore, our framework provides \textbf{physical interpretability and feature attribution} via SHapley Additive exPlanations (SHAP; Figure~\ref{fig:shap_analysis}). Rather than acting as an uninterpretable "black box," the decision tree ensemble re-derives classical gravoturbulent collapse criteria directly from raw simulation data. As shown in Figure~\ref{fig:shap_analysis}a, the model prioritizes physical quantities by placing local gas density ($\text{Density}_i$, $+1.94$), kinetic energy ($\text{Kinetic\_Energy}$, $+1.09$), and magnetic field magnitude ($B_{\rm mag}$, $+0.83$) at the top of the feature hierarchy, while appropriately treating spatial coordinates ($X_i, Y_i, Z_i$) as secondary position flags.

\subsubsection{Limitations}
Despite its success, the model has inherent limitations common to data-driven approximations:

\begin{enumerate}
    \item \textbf{Error Propagation:} As a regression model, errors accumulate over time. While the predictions are robust for the trained horizon ($\Delta T \approx 0.45$ Myr), recursive application of the model (predicting $t_2$ from $t_1$, then $t_3$ from predicted $t_2$) eventually leads to divergence. This is consistent with the chaotic nature of turbulence, where the Lyapunov time limits the horizon of any deterministic prediction.
    
    \item \textbf{Lack of Conservation Laws:} Unlike the MHD solver, XGBoost does not enforce conservation of mass, momentum, or energy. It minimizes statistical error, not physical residuals. While our "Bounded Accuracy" metric shows the results are physically plausible, there is no guarantee that two particles will not be predicted to occupy the same space in unphysical ways (though the pressure term in the training physics indirectly discourages this).

    \item \textbf{Observational Applicability:} The current model relies on full 3D state vectors ($x, y, z, v_x, v_y, v_z, B_x, B_y, B_z$), which are accessible in simulations but not in observations. Observers are limited to 2D projected positions and Line-of-Sight (LOS) velocities. Generalizing this framework to observational data would require retraining the model on "mock observations" (synthetic Position-Position-Velocity cubes), a logical next step for this line of research.

    \item \textbf{Lack of Non-Local Spatial Context:} As demonstrated by our classification task, predicting the absolute fate of a gas parcel (gravitational collapse versus turbulent dispersion) is fundamentally constrained by the point-wise nature of the tabular feature set. Evaluating particles in isolation prevents the model from resolving local spatial differential operators such as magnetic tension $(\mathbf{B} \cdot \nabla)\mathbf{B}$, thermal or magnetic pressure gradients ($\nabla P$), and velocity divergence ($\nabla \cdot \mathbf{v}$), which represent the continuous physical mechanisms required to oppose collapse and establish local force balance in magnetized turbulence.
\end{enumerate}

\subsection{The Role of Magnetic Fields and the Limits of Lagrangian Tracking} \label{sec:mag_fields}

In star formation theory, magnetic fields are known to provide critical pressure support against collapse, regulate accretion rates, and channel gas along field lines. This is reflected in the macroscopic morphological differences between our strongly magnetized ($\beta=0.2$) and moderately magnetized ($\beta=2.0$) validation simulations. 

Incorporating the local magnetic field into our XGBoost framework yielded distinct insights across our two tasks, which are physically elucidated by our feature attribution analysis (Figure~\ref{fig:shap_analysis}). For the continuous regression task, the addition of the local magnetic vectors ($B_x, B_y, B_z$) provided measurable, albeit minor, improvements to the trajectory forecasting. This suggests that while short-term continuous advection is overwhelmingly dominated by bulk gas kinematics and inertia, the machine learning model successfully leverages local magnetic vectors to fine-tune particle paths in strongly magnetized environments.

However, the most profound insight regarding magnetic fields emerged from our binary classification task. By engineering explicit rotational scalars (such as the Alfvénic Mach proxy $v^2/B^2$ and the kinetic-magnetic alignment $\cos \theta$), we achieved an exceptional Average Precision (0.784) and ROC AUC (0.975). The SHAP dependence analysis (Figure~\ref{fig:shap_analysis}b) reveals that the model identifies a sharp non-linear threshold at $\text{Density}_i \approx 0.8$. Below this density boundary, gas parcels receive negative attribution scores ($\text{SHAP} < 0$) and are classified as transient turbulence. Above $\text{Density}_i \approx 0.8$, attribution flips strongly positive, with high magnetic field strengths ($B_{\rm mag} \gtrsim 50$) acting synergistically to yield the highest log-odds attributions for core formation. This physical mechanism is illustrated at the single-particle level in Figure~\ref{fig:shap_analysis}c (Tracer \#21), where high density (+3.13), kinetic energy (+1.85), and magnetic field magnitude (+1.66) drive the output prediction ($f(x) = 7.875$) well above the baseline expectation ($E[f(X)] = -0.016$).

Despite strong global performance, the classifier achieves a limited precision of 37\% at a standard binary decision threshold ($p > 0.5$), systematically over-predicting collapse in boundary cases. Far from being a failure of the machine learning algorithm, this behavior is a direct physical consequence of the point-wise Lagrangian framework. As turbulent gas compresses, magnetic flux freezing causes the local field strength to scale upward alongside density. Evaluating a tracer particle in isolation, the decision tree correctly identifies dense, fast-moving, and strongly magnetized gas as highly susceptible to gravitational collapse. However, the model mathematically lacks access to neighboring gas parcels and therefore cannot calculate spatial derivatives. Without this environmental spatial context, the ensemble cannot determine whether surrounding field line geometry or thermal support is sufficient to counteract gravity and arrest the infall. Consequently, when forced into a binary classification, the model conservatively flags susceptible gas as collapsing, underscoring the physical necessity of spatially connected architectures (e.g., Graph Neural Networks or Neural Operators) to evaluate non-local force balance.

This provides a clear directive for future subgrid modeling. While point-wise Lagrangian methods like XGBoost are exceptionally fast and accurate for advection and probabilistic ranking, creating a truly universal subgrid model that dictates absolute core-formation criteria will require hybrid architectures. Coupling Lagrangian point-tracking with models capable of ingesting local spatial topology, such as Convolutional Neural Networks (CNNs) or Graph Neural Networks (GNNs), will bridge the gap between kinematic advection and continuous force resolution.

%% file: 6_Conclusion.tex
\section{Conclusion} \label{sec:conclusion}

In this paper, we presented a supervised machine learning framework to predict the trajectories and accretion history of gas in turbulent molecular clouds. By leveraging high-resolution ideal MHD simulations and a novel Lagrangian tracer particle pipeline, we transformed the complex, non-linear problem of turbulent core collapse into structured tabular regression and classification tasks.

We successfully trained Extreme Gradient Boosting (XGBoost) models to forecast the 3D positions of gas parcels over a horizon of $\sim 0.31$ Myr ($0.25~t_{\rm ff}$) and to probabilistically classify which parcels would ultimately accrete into prestellar cores. Our key findings are as follows:

\begin{enumerate}
    \item \textbf{Predictive Power of Local Variables:} We demonstrated that the instantaneous localized state (density, velocity, position, and local magnetic field) of a gas parcel contains sufficient information to predict its short-term continuous dynamical evolution with exceptional accuracy ($R^2 > 0.99$).
    
    \item \textbf{Explainable Feature Attribution and Physical Thresholds:} Using SHapley Additive exPlanations (SHAP), we confirmed that the model relies predominantly on physical state variables rather than spatial position artifacts. Local gas density ($\text{Density}_i$, mean $|\text{SHAP}| = +1.94$) dominates the decision hierarchy, followed by kinetic energy ($+1.09$) and magnetic field magnitude ($B_{\rm mag}$, $+0.83$). Crucially, SHAP dependence analysis proved that the model independently re-derived fundamental non-linear MHD collapse criteria—discovering a sharp density transition at $\text{Density}_i \approx 0.8$, above which strong magnetic fields ($B_{\rm mag} \gtrsim 50$) act as a multiplicative factor driving high core-formation probabilities (AUC = 0.975, Average Precision = 0.784).
    
    \item \textbf{Physical Limits of Point-Wise Frameworks:} Despite exceptional global probabilistic rankings ($\text{AUC} = 0.975$), the binary classification task yielded a limited precision of 37\% at standard thresholds. This demonstrates a fundamental physical constraint: while point-wise tabular models are computationally efficient for tracking kinematics, they lack the spatial context required to compute continuous field derivatives. Without access to surrounding neighborhood dynamics to resolve these restoring forces, point-wise models cannot definitively determine whether magnetic support or thermal pressure will halt the collapse of a susceptible density enhancement, underscoring the need for spatially connected architectures to establish absolute core-formation criteria.
\end{enumerate}

This work represents a robust "first-order" approach, utilizing only local, point-wise information. While highly effective for short-term kinematic advection and probabilistic ranking, it inherently cannot resolve the larger morphological context—the filaments, sheets, and extended magnetic field topologies—that drive large-scale flows and provide necessary macroscopic pressure support.

\subsection{Future Work}

The natural physical and computational extension of this work is to incorporate spatial correlations. While XGBoost is state-of-the-art for tabular data, it cannot "see" the geometry of the cloud. Future efforts will leverage Deep Learning architectures, specifically \textit{3D Convolutional Neural Networks (CNNs)} or \textit{Graph Neural Networks (GNNs)}. These models can ingest volumetric grids or local particle neighborhoods, allowing the algorithm to explicitly calculate spatial derivatives and learn from non-local physical mechanisms, such as magnetic tension gradients and filamentary convergence.

By combining the Lagrangian tracking presented here with the spatial context awareness of Deep Learning, we aim to build a unified predictive model for star formation that bridges the gap between turbulent initial conditions and final stellar populations. Furthermore, while this model is primarily designed for simulation subgrid deployment, such purely kinematic frameworks carry significant observational potential. Current observational campaigns, which combine \textit{Gaia} astrometry with high-resolution spectroscopic radial velocities (e.g., APOGEE), are now capable of resolving the requisite 6D phase-space information ($x, y, z, v_x, v_y, v_z$) for young stellar populations. Recent work has demonstrated that these 6D stellar kinematics retain clear physical imprints of the turbulence, gravity, and magnetic fields present within their natal molecular clouds \citep{Velguth2025}. By applying our predictive Lagrangian framework to these high-precision observational datasets, we can begin to bridge the gap between data-driven theoretical models and direct observation, potentially allowing us to reconstruct the dynamic history of star-forming regions directly from the motions of their youngest stars.

%% file: 7_Acknowledgements.tex
\section*{Acknowledgements}

Support for a portion of this work was provided by the National Science Foundation
under Grant AST-1616026.
This work used compute resources provided by the Advanced Cyberinfrastructure Coordination Ecosystem: Services
\& Support (ACCESS) program, which is supported by U.S. National Science
Foundation grants \#2138259, \#2138286, \#2138307, \#2137603, and \#2138296
\citep{Boerner2023}. This work used Anvil at Purdue University \citep{Song2022} through allocation
PHY240036 as well as the Stampede computer at Texas Advanced Computing Center (TACC).

%% file: bibliography.bib
@inproceedings{Boerner2023,
  author    = {Boerner, Timothy J. and Deems, Stephen and Furlani, Thomas R. and Knuth, Shelley L. and Towns, John},
  title     = {ACCESS: Advancing Innovation: NSF's Advanced Cyberinfrastructure Coordination Ecosystem: Services \& Support},
  booktitle = {Practice and Experience in Advanced Research Computing (PEARC '23)},
  year      = {2023},
  pages     = {1--4},
  address   = {Portland, OR, USA},
  publisher = {ACM},
  doi       = {10.1145/3569951.3597559},
  url       = {https://doi.org/10.1145/3569951.3597559}
}

@inproceedings{Song2022,
author = {Song, X. Carol and Smith, Preston and Kalyanam, Rajesh and Zhu, Xiao and Adams, Eric and Colby, Kevin and Finnegan, Patrick and Gough, Erik and Hillery, Elizabett and Irvine, Rick and Maji, Amiya and St. John, Jason},
title = {Anvil - System Architecture and Experiences from Deployment and Early User Operations},
year = {2022},
isbn = {9781450391610},
publisher = {Association for Computing Machinery},
address = {New York, NY, USA},
url = {https://doi.org/10.1145/3491418.3530766},
doi = {10.1145/3491418.3530766},
booktitle = {Practice and Experience in Advanced Research Computing 2022: Revolutionary: Computing, Connections, You},
articleno = {23},
numpages = {9},
location = {Boston, MA, USA},
series = {PEARC '22}
}

@article{Alatalo2013,
   title={The ATLAS3D project – XVIII. CARMA CO imaging survey of early-type galaxies},
   volume={432},
   ISSN={0035-8711},
   url={http://dx.doi.org/10.1093/mnras/sts299},
   DOI={10.1093/mnras/sts299},
   number={3},
   journal={Monthly Notices of the Royal Astronomical Society},
   publisher={Oxford University Press (OUP)},
   author={Alatalo, Katherine and Davis, Timothy A. and Bureau, Martin and Young, Lisa M. and Blitz, Leo and Crocker, Alison F. and Bayet, Estelle and Bois, Maxime and Bournaud, Frédéric and Cappellari, Michele and Davies, Roger L. and de Zeeuw, P. T. and Duc, Pierre-Alain and Emsellem, Eric and Khochfar, Sadegh and Krajnović, Davor and Kuntschner, Harald and Lablanche, Pierre-Yves and Morganti, Raffaella and McDermid, Richard M. and Naab, Thorsten and Oosterloo, Tom and Sarzi, Marc and Scott, Nicholas and Serra, Paolo and Weijmans, Anne-Marie},
   year={2013},
   month=may, pages={1796–1844} }

@ARTICLE{Bate1995,
       author = {{Bate}, Matthew R. and {Bonnell}, Ian A. and {Price}, Nigel M.},
        title = "{Modelling accretion in protobinary systems}",
      journal = {\mnras},
         year = 1995,
        month = nov,
       volume = {277},
       number = {2},
        pages = {362-376},
          doi = {10.1093/mnras/277.2.362},
archivePrefix = {arXiv},
       eprint = {astro-ph/9510149},
 primaryClass = {astro-ph},
       adsurl = {https://ui.adsabs.harvard.edu/abs/1995MNRAS.277..362B}
}

@article{Bryan2014,
   title={ENZO: AN ADAPTIVE MESH REFINEMENT CODE FOR ASTROPHYSICS},
   volume={211},
   ISSN={1538-4365},
   url={http://dx.doi.org/10.1088/0067-0049/211/2/19},
   DOI={10.1088/0067-0049/211/2/19},
   number={2},
   journal={The Astrophysical Journal Supplement Series},
   publisher={American Astronomical Society},
   author={Bryan, Greg L. and Norman, Michael L. and O’Shea, Brian W. and Abel, Tom and Wise, John H. and Turk, Matthew J. and Reynolds, Daniel R. and Collins, David C. and Wang, Peng and Skillman, Samuel W. and Smith, Britton and Harkness, Robert P. and Bordner, James and Kim, Ji-hoon and Kuhlen, Michael and Xu, Hao and Goldbaum, Nathan and Hummels, Cameron and Kritsuk, Alexei G. and Tasker, Elizabeth and Skory, Stephen and Simpson, Christine M. and Hahn, Oliver and Oishi, Jeffrey S. and So, Geoffrey C. and Zhao, Fen and Cen, Renyue and Li, Yuan},
   year={2014},
   month=mar, pages={19} }

@inproceedings{Chen2016, series={KDD ’16},
   title={XGBoost: A Scalable Tree Boosting System},
   url={http://dx.doi.org/10.1145/2939672.2939785},
   DOI={10.1145/2939672.2939785},
   booktitle={Proceedings of the 22nd ACM SIGKDD International Conference on Knowledge Discovery and Data Mining},
   publisher={ACM},
   author={Chen, Tianqi and Guestrin, Carlos},
   year={2016},
   month=aug, pages={785–794},
   collection={KDD ’16} }

@misc{Chevance2022,
      title={The Life and Times of Giant Molecular Clouds}, 
      author={Mélanie Chevance and Mark R. Krumholz and Anna F. McLeod and Eve C. Ostriker and Erik W. Rosolowsky and Amiel Sternberg},
      year={2022},
      eprint={2203.09570},
      archivePrefix={arXiv},
      primaryClass={astro-ph.GA}
}

@ARTICLE{Collins2010,
   author = {{Collins}, D.~C. and {Xu}, H. and {Norman}, M.~L. and {Li}, H. and 
    {Li}, S.},
    title = "{Cosmological Adaptive Mesh Refinement Magnetohydrodynamics with Enzo}",
  journal = {\apjs},
archivePrefix = "arXiv",
   eprint = {0902.2594},
     year = 2010,
    month = feb,
   volume = 186,
    pages = {308-333},
      doi = {10.1088/0067-0049/186/2/308},
   adsurl = {http://adsabs.harvard.edu/abs/2010ApJS..186..308C}
}

@article{Collins2012,
   title={THE TWO STATES OF STAR-FORMING CLOUDS},
   volume={750},
   ISSN={1538-4357},
   url={http://dx.doi.org/10.1088/0004-637X/750/1/13},
   DOI={10.1088/0004-637x/750/1/13},
   number={1},
   journal={The Astrophysical Journal},
   publisher={American Astronomical Society},
   author={Collins, David C. and Kritsuk, Alexei G. and Padoan, Paolo and Li, Hui and Xu, Hao and Ustyugov, Sergey D. and Norman, Michael L.},
   year={2012},
   month=apr, pages={13} }

@ARTICLE{Collins2023,
       author = {{Collins}, David C. and {Le}, Dan and {Jimenez Vela}, Luz L.},
        title = "{Collapsing molecular clouds with tracer particles - I. What collapses?}",
      journal = {\mnras},
         year = 2023,
        month = apr,
       volume = {520},
       number = {3},
        pages = {4194-4208},
          doi = {10.1093/mnras/stac2834},
       adsurl = {https://ui.adsabs.harvard.edu/abs/2023MNRAS.520.4194C}
}

@misc{Collins2024,
      title={Collapsing molecular clouds with tracer particles: Part II, Collapse Histories}, 
      author={David C. Collins and Dan K. Le and Luz L. Jimenez Vela},
      year={2024},
      eprint={2306.10320},
      archivePrefix={arXiv},
      primaryClass={astro-ph.GA},
      url={https://arxiv.org/abs/2306.10320}, 
}

@ARTICLE{Crutcher2012,
       author = {{Crutcher}, Richard M.},
        title = "{Magnetic Fields in Molecular Clouds}",
      journal = {araa},
         year = 2012,
        month = sep,
       volume = {50},
        pages = {29-63},
          doi = {10.1146/annurev-astro-081811-125514},
       adsurl = {https://ui.adsabs.harvard.edu/abs/2012ARA&A..50...29C}
}

@article{Dieleman2015,
   title={Rotation-invariant convolutional neural networks for galaxy morphology prediction},
   volume={450},
   ISSN={0035-8711},
   url={http://dx.doi.org/10.1093/mnras/stv632},
   DOI={10.1093/mnras/stv632},
   number={2},
   journal={Monthly Notices of the Royal Astronomical Society},
   publisher={Oxford University Press (OUP)},
   author={Dieleman, Sander and Willett, Kyle W. and Dambre, Joni},
   year={2015},
   month=apr, pages={1441–1459} }

@ARTICLE{Evans2009,
       author = {{Evans}, II, Neal J. and {Dunham}, Michael M. and {J{\o}rgensen}, Jes K. and {Enoch}, Melissa L. and {Mer{\'\i}n}, Bruno and {van Dishoeck}, Ewine F. and {Alcal{\'a}}, Juan M. and {Myers}, Philip C. and {Stapelfeldt}, Karl R. and {Huard}, Tracy L. and {Allen}, Lori E. and {Harvey}, Paul M. and {van Kempen}, Tim and {Blake}, Geoffrey A. and {Koerner}, David W. and {Mundy}, Lee G. and {Padgett}, Deborah L. and {Sargent}, Anneila I.},
        title = "{The Spitzer c2d Legacy Results: Star-Formation Rates and Efficiencies; Evolution and Lifetimes}",
      journal = {\apjs},
         year = 2009,
        month = apr,
       volume = {181},
       number = {2},
        pages = {321-350},
          doi = {10.1088/0067-0049/181/2/321},
archivePrefix = {arXiv},
       eprint = {0811.1059},
 primaryClass = {astro-ph},
       adsurl = {https://ui.adsabs.harvard.edu/abs/2009ApJS..181..321E}
}

@article{Federrath2010b,
	author = {{Federrath}, Christoph and {Banerjee}, Robi and {Clark}, Paul C. and {Klessen}, Ralf S.},
	title = {{Modeling Collapse and Accretion in Turbulent Gas Clouds: Implementation and Comparison of Sink Particles in AMR and SPH}},
	journal = {\apj},
	year = {2010},
	month = {apr},
	volume = {713},
	number = {1},
	pages = {269-290},
	doi = {10.1088/0004-637X/713/1/269},
	archivePrefix = {arXiv},
	eprint = {1001.4456},
	primaryClass = {astro-ph.SR},
	adsurl = {https://ui.adsabs.harvard.edu/abs/2010ApJ...713..269F}
}

@ARTICLE{Federrath2012,
       author = {{Federrath}, Christoph and {Klessen}, Ralf S.},
        title = "{The Star Formation Rate of Turbulent Magnetized Clouds: Comparing Theory, Simulations, and Observations}",
      journal = {\apj},
         year = 2012,
        month = dec,
       volume = {761},
       number = {2},
          eid = {156},
        pages = {156},
          doi = {10.1088/0004-637X/761/2/156},
archivePrefix = {arXiv},
       eprint = {1209.2856},
 primaryClass = {astro-ph.SR},
       adsurl = {https://ui.adsabs.harvard.edu/abs/2012ApJ...761..156F}
}

@ARTICLE{Fryxell2000,
       author = {{Fryxell}, B. and {Olson}, K. and {Ricker}, P. and {Timmes}, F.~X. and {Zingale}, M. and {Lamb}, D.~Q. and {MacNeice}, P. and {Rosner}, R. and {Truran}, J.~W. and {Tufo}, H.},
        title = "{FLASH: An Adaptive Mesh Hydrodynamics Code for Modeling Astrophysical Thermonuclear Flashes}",
      journal = {\apjs},
         year = 2000,
        month = nov,
       volume = {131},
       number = {1},
        pages = {273-334},
          doi = {10.1086/317361},
       adsurl = {https://ui.adsabs.harvard.edu/abs/2000ApJS..131..273F}
}

@article{Gardiner2005,
	author = {{Gardiner}, T.~A. and {Stone}, J.~M.},
	title = {{An unsplit Godunov method for ideal MHD via constrained transport}},
	journal = {\jcompphys},
	eprint = {arXiv:astro-ph/0501557},
	year = {2005},
	month = {may},
	volume = {205},
	pages = {509-539},
	doi = {10.1016/j.jcp.2004.11.016},
	adsurl = {http://adsabs.harvard.edu/abs/2005JCoPh.205..509G}
}

@ARTICLE{Krumholz2005,
       author = {{Krumholz}, Mark R. and {McKee}, Christopher F.},
        title = "{A General Theory of Turbulence-regulated Star Formation, from Spirals to Ultraluminous Infrared Galaxies}",
      journal = {apj},
         year = 2005,
        month = sep,
       volume = {630},
       number = {1},
        pages = {250-268},
          doi = {10.1086/431734},
archivePrefix = {arXiv},
       eprint = {astro-ph/0505177},
 primaryClass = {astro-ph},
       adsurl = {https://ui.adsabs.harvard.edu/abs/2005ApJ...630..250K}
}

@ARTICLE{Krumholz2007,
       author = {{Krumholz}, Mark R. and {Tan}, Jonathan C.},
        title = "{Slow Star Formation in Dense Gas: Evidence and Implications}",
      journal = {\apj},
         year = 2007,
        month = jan,
       volume = {654},
       number = {1},
        pages = {304-315},
          doi = {10.1086/509101},
archivePrefix = {arXiv},
       eprint = {astro-ph/0606277},
 primaryClass = {astro-ph},
       adsurl = {https://ui.adsabs.harvard.edu/abs/2007ApJ...654..304K}
}

@ARTICLE{Larson1981,
       author = {{Larson}, R.~B.},
        title = "{Turbulence and star formation in molecular clouds.}",
      journal = {mnras},
         year = 1981,
        month = mar,
       volume = {194},
        pages = {809-826},
          doi = {10.1093/mnras/194.4.809},
       adsurl = {https://ui.adsabs.harvard.edu/abs/1981MNRAS.194..809L}
}

@article{Li2008,
	author = {{Li}, S. and {Li}, H. and {Cen}, R.},
	title = {{CosmoMHD: A Cosmological Magnetohydrodynamics Code}},
	journal = {\apjs},
	year = {2008},
	month = {jan},
	volume = {174},
	pages = {1-12},
	doi = {10.1086/521302},
	adsurl = {http://adsabs.harvard.edu/abs/2008ApJS..174....1L}
}

@misc{Lundberg2017,
      title={A Unified Approach to Interpreting Model Predictions}, 
      author={Scott Lundberg and Su-In Lee},
      year={2017},
      eprint={1705.07874},
      archivePrefix={arXiv},
      primaryClass={cs.AI},
      url={https://arxiv.org/abs/1705.07874}, 
}

@article{Lundberg2020,
  author    = {Lundberg, Scott M. and Erion, Gabriel and Chen, Hugh and DeGrave, Alex and Prutkin, Jordan M. and Nair, Bala and Katz, Ronit and Himmelfarb, Jonathan and Bansal, Nisha and Lee, Su-In},
  title     = {From Local Explanations to Global Understanding with Explainable {AI} for Trees},
  journal   = {Nature Machine Intelligence},
  volume    = {2},
  number    = {1},
  pages     = {56--67},
  year      = {2020},
  month     = {jan},
  doi       = {10.1038/s42256-019-0138-9},
  note      = {PMID: 32607472; PMCID: PMC7326367}
}

@article{MacLow1999,
   title={The Energy Dissipation Rate of Supersonic, Magnetohydrodynamic Turbulence in Molecular Clouds},
   volume={524},
   ISSN={1538-4357},
   url={http://dx.doi.org/10.1086/307784},
   DOI={10.1086/307784},
   number={1},
   journal={The Astrophysical Journal},
   publisher={American Astronomical Society},
   author={Mac Low, Mordecai‐Mark},
   year={1999},
   month=oct, pages={169–178} }

@ARTICLE{MacLow2004,
       author = {{Mac Low}, Mordecai-Mark and {Klessen}, Ralf S.},
        title = "{Control of star formation by supersonic turbulence}",
      journal = {Reviews of Modern Physics},
         year = 2004,
        month = jan,
       volume = {76},
       number = {1},
        pages = {125-194},
          doi = {10.1103/RevModPhys.76.125},
archivePrefix = {arXiv},
       eprint = {astro-ph/0301093},
 primaryClass = {astro-ph},
       adsurl = {https://ui.adsabs.harvard.edu/abs/2004RvMP...76..125M}
}

@ARTICLE{McKee2007,
       author = {{McKee}, Christopher F. and {Ostriker}, Eve C.},
        title = "{Theory of Star Formation}",
      journal = {araa},
         year = 2007,
        month = sep,
       volume = {45},
       number = {1},
        pages = {565-687},
          doi = {10.1146/annurev.astro.45.051806.110602},
archivePrefix = {arXiv},
       eprint = {0707.3514},
 primaryClass = {astro-ph},
       adsurl = {https://ui.adsabs.harvard.edu/abs/2007ARA&A..45..565M}
}

@article{Mignone2007,
	author = {{Mignone}, A.},
	title = {{A simple and accurate Riemann solver for isothermal MHD}},
	journal = {\jcompphys},
	eprint = {arXiv:astro-ph/0701798},
	year = {2007},
	month = {aug},
	volume = {225},
	pages = {1427-1441},
	doi = {10.1016/j.jcp.2007.01.033},
	adsurl = {http://adsabs.harvard.edu/abs/2007JCoPh.225.1427M}
}

@ARTICLE{Ntampaka2015,
       author = {{Ntampaka}, M. and {Trac}, H. and {Sutherland}, D.~J. and {Battaglia}, N. and {P{\'o}czos}, B. and {Schneider}, J.},
        title = "{A Machine Learning Approach for Dynamical Mass Measurements of Galaxy Clusters}",
      journal = {\apj},
         year = 2015,
        month = apr,
       volume = {803},
       number = {2},
          eid = {50},
        pages = {50},
          doi = {10.1088/0004-637X/803/2/50},
archivePrefix = {arXiv},
       eprint = {1410.0686},
 primaryClass = {astro-ph.CO},
       adsurl = {https://ui.adsabs.harvard.edu/abs/2015ApJ...803...50N}
}

@ARTICLE{Padoan2002,
       author = {{Padoan}, Paolo and {Nordlund}, {\r{A}}ke},
        title = "{The Stellar Initial Mass Function from Turbulent Fragmentation}",
      journal = {\apj},
         year = 2002,
        month = sep,
       volume = {576},
       number = {2},
        pages = {870-879},
          doi = {10.1086/341790},
archivePrefix = {arXiv},
       eprint = {astro-ph/0011465},
 primaryClass = {astro-ph},
       adsurl = {https://ui.adsabs.harvard.edu/abs/2002ApJ...576..870P}
}

@ARTICLE{Peek2019,
       author = {{Peek}, J.~E.~G. and {Burkhart}, Blakesley},
        title = "{Do Androids Dream of Magnetic Fields? Using Neural Networks to Interpret the Turbulent Interstellar Medium}",
      journal = {\apjl},
         year = 2019,
        month = sep,
       volume = {882},
       number = {1},
          eid = {L12},
        pages = {L12},
          doi = {10.3847/2041-8213/ab3a9e},
archivePrefix = {arXiv},
       eprint = {1905.00918},
 primaryClass = {astro-ph.IM},
       adsurl = {https://ui.adsabs.harvard.edu/abs/2019ApJ...882L..12P}
}

@article{Rosolowsky2008,
	author = {{Rosolowsky}, E.~W. and {Pineda}, J.~E. and {Kauffmann}, J. and 
   {Goodman}, A.~A.},
	title = {{Structural Analysis of Molecular Clouds: Dendrograms}},
	journal = {\apj},
	archivePrefix = {arXiv},
	eprint = {0802.2944},
	year = {2008},
	month = {jun},
	volume = {679},
	pages = {1338-1351},
	doi = {10.1086/587685},
	adsurl = {http://adsabs.harvard.edu/abs/2008ApJ...679.1338R}
}

@ARTICLE{Sahu2021,
       author = {{Sahu}, Dipen and {Liu}, Sheng-Yuan and {Liu}, Tie},
        title = "{Anatomy of Orion molecular clouds - the astrochemistry perspective/approach}",
      journal = {Frontiers in Astronomy and Space Sciences},
         year = 2021,
        month = oct,
       volume = {8},
          eid = {137},
        pages = {137},
          doi = {10.3389/fspas.2021.672893},
       adsurl = {https://ui.adsabs.harvard.edu/abs/2021FrASS...8..137S}
}

@misc{Salim2024,
      title={Extending a Physics-Informed Machine Learning Network for Superresolution Studies of Rayleigh-B\'enard Convection}, 
      author={Diane M. Salim and Blakesley Burkhart and David Sondak},
      year={2024},
      eprint={2307.02674},
      archivePrefix={arXiv},
      primaryClass={physics.flu-dyn},
      url={https://arxiv.org/abs/2307.02674}, 
}

@misc{Schmaltz2025,
      title={Estimate Sonic Mach Number in the Interstellar Medium with Convolutional Neural Network}, 
      author={Tyler Schmaltz and Yue Hu and Alex Lazarian},
      year={2025},
      eprint={2411.11157},
      archivePrefix={arXiv},
      primaryClass={astro-ph.GA},
      url={https://arxiv.org/abs/2411.11157}, 
}

@inbook{Shapley2016,
url = {https://doi.org/10.1515/9781400881970-018},
title = {17. A Value for n-Person Games},
booktitle = {Contributions to the Theory of Games, Volume II},
author = {L. S. Shapley},
editor = {Harold William Kuhn and Albert William Tucker},
publisher = {Princeton University Press},
address = {Princeton},
pages = {307--318},
doi = {doi:10.1515/9781400881970-018},
isbn = {9781400881970},
year = {2016},
lastchecked = {2026-08-24}
}

@ARTICLE{Shu1977,
       author = {{Shu}, F.~H.},
        title = "{Self-similar collapse of isothermal spheres and star formation.}",
      journal = {apj},
         year = 1977,
        month = jun,
       volume = {214},
        pages = {488-497},
          doi = {10.1086/155274},
       adsurl = {https://ui.adsabs.harvard.edu/abs/1977ApJ...214..488S}
}

@article{Stone1974,
    author = {Stone, M.},
    title = {Cross-Validatory Choice and Assessment of Statistical Predictions},
    journal = {Journal of the Royal Statistical Society: Series B (Methodological)},
    volume = {36},
    number = {2},
    pages = {111-133},
    year = {1974},
    month = {12},
    issn = {0035-9246},
    doi = {10.1111/j.2517-6161.1974.tb00994.x},
    url = {https://doi.org/10.1111/j.2517-6161.1974.tb00994.x},
    eprint = {https://academic.oup.com/jrsssb/article-pdf/36/2/111/49096683/jrsssb\_36\_2\_111.pdf},
}

@ARTICLE{Tan2024,
       author = {{Tan}, Lei and {Liu}, Zhicun and {Wang}, Xiaolong and {Mei}, Ying and {Wang}, Feng and {Deng}, Hui and {Liu}, Chao},
        title = "{A Robust Young Stellar Object Identification Method Based on Deep Learning}",
      journal = {\apjs},
         year = 2024,
        month = aug,
       volume = {273},
       number = {2},
          eid = {34},
        pages = {34},
          doi = {10.3847/1538-4365/ad5a08},
       adsurl = {https://ui.adsabs.harvard.edu/abs/2024ApJS..273...34T}
}

@misc{Velguth2025,
      title={Probing Turbulence, Gravity, Supernovae, and Magnetic Field Effects with the 6D Kinematics of Young Stars in Milky Way Star-Forming Regions}, 
      author={Benjamin N. Velguth and Yuan Li and Trung Ha and Yue Hu and Marina Kounkel and Siyao Xu and Robert Gutermuth and Hui Li and Martin D. Weinberg},
      year={2025},
      eprint={2507.22107},
      archivePrefix={arXiv},
      primaryClass={astro-ph.GA},
      url={https://arxiv.org/abs/2507.22107}, 
}

@article{Wang2009,
   title={OUTFLOW FEEDBACK REGULATED MASSIVE STAR FORMATION IN PARSEC-SCALE CLUSTER-FORMING CLUMPS},
   volume={709},
   ISSN={1538-4357},
   url={http://dx.doi.org/10.1088/0004-637X/709/1/27},
   DOI={10.1088/0004-637x/709/1/27},
   number={1},
   journal={The Astrophysical Journal},
   publisher={American Astronomical Society},
   author={Wang, Peng and Li, Zhi-Yun and Abel, Tom and Nakamura, Fumitaka},
   year={2009},
   month=dec, pages={27–41} }

@article{Weinberger2020,
   title={The AREPO Public Code Release},
   volume={248},
   ISSN={1538-4365},
   url={http://dx.doi.org/10.3847/1538-4365/ab908c},
   DOI={10.3847/1538-4365/ab908c},
   number={2},
   journal={The Astrophysical Journal Supplement Series},
   publisher={American Astronomical Society},
   author={Weinberger, Rainer and Springel, Volker and Pakmor, Rüdiger},
   year={2020},
   month=jun, pages={32} }

@misc{Zhang2026,
      title={Machine-Learning Estimation of Energy Fractions in MHD Turbulence Modes}, 
      author={Jiyao Zhang and Yue Hu},
      year={2026},
      eprint={2511.04119},
      archivePrefix={arXiv},
      primaryClass={astro-ph.GA},
      url={https://arxiv.org/abs/2511.04119}, 
}
